\documentclass[pra,aps,twocolumn,reprint,floatfix,10pt]{revtex4-1}
\usepackage{graphicx}
\usepackage{amsmath} 
\usepackage[T1]{fontenc}
\usepackage{mathtools}
\usepackage[colorlinks,citecolor=blue,linkcolor=magenta,urlcolor=red]{hyperref}

\def\be{\begin{equation}}
\def\ee{\end{equation}}
\def\bea{\begin{eqnarray}}
\def\eea{\end{eqnarray}}
\usepackage{ulem}
\def\IPR{\mathrm{IPR}}

\begin{document}
\title{Single-particle localization in dynamical potentials}
\author{Jan Major} 
\affiliation{
Instytut Fizyki imienia Mariana Smoluchowskiego, 
Uniwersytet Jagiello\'nski, ulica Profesora Stanis\l{}awa \L{}ojasiewicza 11, PL-30-348 Krak\'ow, Poland}
\author{Giovanna Morigi}
\affiliation{Theoretische Physik, Universit\"at des Saarlandes, D-66123  Saarbr\"ucken, Germany}
\author{Jakub Zakrzewski} 
\affiliation{
Instytut Fizyki imienia Mariana Smoluchowskiego, 
Uniwersytet Jagiello\'nski, ulica Profesora Stanis\l{}awa \L{}ojasiewicza 11, PL-30-348 Krak\'ow, Poland}
\affiliation{Mark Kac Complex Systems Research Center, Uniwersytet Jagiello\'nski, ulica Profesora Stanis\l{}awa \L{}ojasiewicza 11, PL-30-348 Krak\'ow, Poland}

\begin{abstract}
Single particle localization of an ultra-cold atom {is studied in one dimension when the atom is confined by an optical lattice and by the incommensurate potential of a high-finesse optical cavity. In the strong coupling regime the atom is a dynamical refractive medium, the cavity resonance depends on the atomic position within the standing-wave mode and nonlinearly determines the depth and form of the incommensurate potential.  We show that the particular form of the quasi-random cavity potential leads to the appearance of mobility edges, even in presence of nearest-neighbour hopping. We provide a detailed characterization of the system as a function of its  parameters and in particular of the strength of the atom-cavity coupling, which {controls the functional form of the cavity potential}. For strong atom-photon coupling the properties of the mobility edges significantly depend on the ratio between the periodicities of the confining optical lattice and of the cavity field.}
\end{abstract}
\maketitle
\section{Introduction}

The Aubry-Andr\'e model \cite{Aubry80} describes a quantum particle tightly confined by a one-dimensional lattice and in presence of a second periodic potential, a harmonic function whose period is incommensurate with the main lattice period.
In such a system, 
when the second potential exceeds some critical heigth all states are exponentially localized, much likely as in the case of Anderson localization in truly disordered systems \cite{Anderson58}. The fact that {localization indeed manifests} in the Aubry-Andr\'e model have been formally proven in Ref. \cite{Jitomirskaya99}. Due to its 
spatial correlations the Aubry-Andr\'e potential and its extensions are often referred to as \emph{quasi-disordered} potentials.

Detailed studies of disorder-induced effects is presently possible since ultracold atomic systems allow unprecedented level of controllability over the system parameters \cite{Bloch08, Jaksch2005,Lewenstein12}. This is particularly true for optical lattice potentials where  different lattice geometries can be realised \cite{Windpassinger13}, on-site potentials as well as tunnelings can be tailored. 
Moreover, artificial gauge fields can be simulated often adapting periodic modulations of lattice parameters or interactions \cite{holthaus2005,Lignier2007,Sengstock12,Targonska12,Eckardt10,Rapp12,przysiezna2015,Dutta2015,Major17,Dutta17}. Of particular value is the control over the interaction strength by means of Feshbach resonances \cite{Chin10}.

Early propositions to study disorder-induced effects in cold atom settings \cite{Damski03,Roth03} {soon resulted} in experimental attempts to observe direct signatures of localization in interacting condensates \cite{Lye05,Clement05,Fort05,Schulte05,Schulte06}. Only when interactions were turned off localization could be directly observed in ultracold atomic gases placed in speckle \cite{Billy08} or quasi-random \cite{Roati08} potentials in quasi one-dimensional (1D) system. The latter case is precisely the case of Aubry-Andr\'e localization \cite{Aubry80}.
Soon afterwards further progress was made leading to the demonstration of three-dimensional (3D) Anderson localization \cite{Kondov11,Jendrzejewski12}. With the development of many-body localization theory \cite{Basko06,Oganesyan07} it became clear that a sufficiently strong disorder leads also to localization for interacting particles breaking a common wisdom of ergodicity in such systems \cite{Srednicki94}. Research on many-body localization has rapidly advanced in recent years (see e.g. reviews \cite{Huse14,Rahul15} ) followed by exciting experimental developments
\cite{Schreiber15,Choi16}. While early theory work considers spin systems (reducing in some cases to spinless fermions) many-body localization is predicted to occur also for bosons
\cite{Sierant17,Sierant17b,Sierant18}. Let us also note that Anderson localization is predicted to occur for solitons, namely, even for weak disorder and in the presence of interactions \cite{Sacha09}.

Notwithstanding the rising interesting on localization in interacting systems, the non-interacting limit is still at the center of intensive studies on the critical dynamics close to the localization transition
\cite{Evers08,Mueller16}.  Moreover, the position of the mobility edge for 3D Anderson localization is subject of current debate \cite{Semeghini15,Pasek17}. Anderson localization is studied in a variety of systems, recent propositions suggest that the phenomenon can occur in the time domain \cite{Sacha16,Delande17} and it can be understood in terms of time crystals \cite{Wilczek12,Sacha15,Giergiel18}
(for a review see \cite{Sacha18}). The one-dimensional case presents peculiar features. Here, even a tiny truly random disorder leads to localization of all eigenstates. On the contrary, Aubry-Andr\'e localization in quasi-periodic potentials occurs at a threshold value. 
This behaviour is modified when hopping in the main lattice has tails beyond nearest-neighbour coupling \cite{Biddle11,Deng18}. In this modified Aubry-Andr\'e model one may observe (as in the standard 3D case) mobility edges, i.e. situations where, for a given disorder, the energy eigenstates within a band can be delocalized or localized and are separated in energy by a ``mobility edge'' \cite{Lueschen18}. Similarly nontrivial correlations in disorder  or nondiagonal disorder (e.g. random tunnelings) lead to an appearance of mobility edges \cite{Flores89,Izrailev01,Peng04,Kogan08,Schaff10,Plodzien11,Wang13,Kosior15,Kosior15b,Liu16,Major16}.

In this work we show that mobility edges with peculiar features can appear in a different extension of Aubry-Andr\'e model, where the hopping is nearest-neighbour and uniform while the incommensurate potential, in turn, is not a simple harmonic function but instead can possess all higher harmonics. This model is an idealization of the dynamics of atoms which are confined by optical lattice potentials and interacts with a standing-wave mode of a high-finesse optical resonator {when the system is pumped by a laser which either couple directly to the cavity or pumps transversally the atoms. Here, the strong optomechanical coupling with the atoms gives rise to a shift of the cavity resonance which depends on the atomic density within the cavity standing wave, and thus to a nonlinear dependence of the intracavity potential on the atomic density \cite{Larson08,Baumann10,Fernandez10,Ritsch13}. When the atoms are transversally pumped by the laser and} the periodicity of cavity mode and optical lattice are commensurate, interacting atoms can form density-wave phases \cite{Klinder15,Landig16,Dogra16,Niederle16}. Similarly, the incommensurate ratio of these frequencies may lead to a quasi-random potential, giving rise to disordered phases in interacting 
systems \cite{Habibian13,Habibian13b}. The ground state of a single cold atom for incommensurate ratios was analysed in Ref. \cite{Rojan16}. {It was shown that 
the specific incommensurate potential of the cavity field -- compare Eq.~\eqref{onsite} below -- }leads to Anderson-like localization of the ground state. This localization is due to the incommensurate potential which emerges because of cavity back-action and is thus self-induced by the atom. It has been argued that the dynamics of atomic wave packets in a related model can exhibit anomalous diffusion \cite{Zheng18}.
In this work we {significantly extend the previous study of Ref. \cite{Rojan16} by analysing the properties of excited states in the configuration originally proposed in \cite{Rojan16}. }We  show that this model can exhibit a mobility edge. The appearance of a mobility edge depends on the strength of the coupling between the atom and the cavity mode, and results thus from { the nonlinear character  of the optomechanical potential. Interestingly the system's behavior exhibits a dramatic dependence on the incommensurability parameter, and it is thus sensitive to the quasi-random disorder} {of the self-induced cavity potential.}
The paper is structured as follows. In section \ref{sec:Model} we present the model used and we sketch its derivation. further in section \ref{sec:Results} we show and discuss the results of the numerical calculations. The conclusions are drawn in section \ref{sec:Conclusions}.

\section{Model\label{sec:Model}}

In this section we introduce and justify the model which is the starting point of our investigation. The material here presented summarizes the detailed derivations reported in Refs. \cite{Fernandez10,Habibian13b,Rojan16}

\subsection{Optomechanical coupling between atomic motion and cavity}

The system we consider is an atom of mass $m$ whose motion is constrained along one dimension, which we identify here with the $x$-axis. The atomic motion is tightly bound by an optical lattice and confined inside a high-finesse optical resonator, which in turn is driven by a laser.
An atomic dipolar transition strongly couples to one standing-wave mode of the resonator which dissipates photons at rate $\kappa$. We consider the limit in which the coupling is purely optomechanical, namely, the atomic internal degrees of freedom can be described by the dispersive polarizability and cavity and atomic motion are directly coupled to one another. The Hamiltonian part of the dynamics takes the form
\begin{align}
\label{H:opto}
H_\mathrm{opto}=&\frac{p^2}{2m}+W_0 \cos^2 (2\pi x/\lambda_0)-\hbar \Delta_c a^\dagger a\\
& +{\rm i}\hbar \eta (a^\dagger-a)+\hbar U_0a^\dagger a \cos (2\pi x/\lambda)\,,\nonumber
\end{align}
where $p$ and $x$ are the canonically-conjugated momentum and position of the atom, $W_0$ and $\lambda_0$ are the depth and wavelength of the optical lattice, respectively, $a$ and $a^\dagger$ are the annihilation and creation operator of a cavity photon at frequency $\omega_c$ and wavelength $\lambda$. Parameter $\eta$ denote the strength of the pumping laser at frequency $\omega_p$ and the Hamiltonian is reported in the reference frame rotating at the pump frequency, with $\Delta_c=\omega_p-\omega_c$  the detuning between pump and cavity frequency. Finally, the optomechanical coupling between cavity and atomic motion gives rise to an optical lattice at periodicity $\lambda$ and depth $U_0 a^\dagger a$. In turn, this term also describes a shift of the cavity frequency which depends on the atomic position, $U_0  \cos (2\pi x/\lambda)$. Therefore, this term gives rise to a nonlinear coupling between atomic motion and resonator which is scaled by the parameter $U_0$. In particular, $U_0$ can be either positive 
or negative depending on the sign of the atomic detuning \cite{Ritsch13}.
  
The incoherent part of the dynamics is solely given by cavity losses and is described by a Born-Markov master equation for the density matrix $\rho$ of cavity and atom's external degrees of freedom. The full master equation reads 
\begin{align}
\partial_t \rho=\frac{1}{i\hbar}[H_\mathrm{opto},\rho]+\mathcal{L}\rho\,,
\end{align}
where dissipator $\mathcal{L}$ describes the cavity losses:
\begin{align}
 \mathcal{L}\rho=\kappa (2a\rho a^\dagger-a^\dagger a \rho-\rho a^\dagger a)\,.
\end{align}
Losses due to spontaneous emission are here neglected, since the fields are far-off resonance from the optical dipole transition so that the atom-light interactions are in the dispersive regime. 

In the rest of this work we will focus on the atomic motion. This is confined by two potentials: the external optical lattice, at fixed depth, and the cavity standing-wave potential, whose depth is proportional to the number of intracavity photons and is thus a dynamical variable. We focus on the dynamics when the ratio $\beta$ between the two periodicity, $\beta=\lambda_0/\lambda$, is incommensurate. This situation would reproduce the Aubry-Andr\'e model, but with an important difference due to the optomechanical coupling, which gives rise to an effective nonlinearity in the dynamics of the atomic motion for sufficiently large values of $U_0$. Parameter $U_0$, indeed, is related to the dispersive cooperativity $C_0$ of cavity quantum electrodynamics \cite{Kimble94} by the relation $C_0=|U_0|/\kappa$. In the regime where $C_0\ge 1$ the backaction of the cavity field on the atomic motion thus appears in terms of a potential which contains higher harmonics than the one at wave number $k=2\pi/\lambda$. This 
becomes evident in the limit in which the cavity degrees of freedom can be eliminated from the equations of motion of the atom, which is regime on which we focus on the rest of this paper. 

\subsection{Eliminating the cavity degrees of freedom}

We now consider the regime in which the characteristic time scale $\tau_c$ of the cavity degrees of freedom is orders of magnitude smaller than characteristic time scale $T_M$ of the atomic motion. {In this regime $\tau_c\sim |\Delta_c+{\rm i}\kappa|^{-1}$ and $T_M\sim \sqrt{\omega_R E_{\rm kin}}$, where $\omega_R=2\pi^2/(m\lambda^2)$ is the recoil energy and $E_{\rm kin}$ is the average atom's kinetic energy. 
For $\sqrt{\kappa^2+\Delta_c^2}\gg \sqrt{\omega_R E_{\rm kin}}$ we can perform} a coarse graining over the time $\Delta t$, such that $\tau_c\ll\Delta t\ll T_M$. Moreover, for sufficiently large ratios $T_M/\tau_c$ the cavity shot noise can be neglected and the field variable can be replaced by its average value over $\Delta t$, which is now a function of the atomic variable: $$a\to \bar a \approx \eta/[(\Delta_c-U_0\cos^2(k x))+{\rm i}\kappa]\,.$$ Details of this procedure can be found in Refs. \cite{Larson08,Fernandez10,Habibian13} (see Ref.~\cite{Schuetz13} for the semiclassical approximation). In this limit the dynamics is described by the effective Hamiltonian:
\begin{align}
\label{eq:H0}
H_\mathrm{eff}=\frac{p^2}{2m}+W_0 \cos^2 (\pi x)+\varepsilon(x)\,,
\end{align}
where we have now reported the position $x$ in units of $\lambda_0/2$. The term $\varepsilon(x)$ is the nonlinear potential due to cavity backaction, which reads \cite{Rojan16}
$$\varepsilon(x) = V_0 \mathrm{arctan}(-\delta+C\cos^2(\beta\pi x))\,,$$
with $V_0=\hbar\eta^2/\kappa$ a proportionality factor whose strength is controlled by the pump intensity, $\delta=\Delta_c/\kappa$, $C=U_0/\kappa$ (such that $C_0=|C|$), and $\beta$ some irrational number. In the first part of this paper we set it equal to the golden ratio $\phi=(1+\sqrt{5})/2$. 

The parameter $C$, which we will denote from now on by "cooperativity", can be both negative and positive, depending on the sign of $U_0$ and thus of the detuning between fields and atomic transition. This is important for the following discussion, since when $C>0$ the minima of the cavity potential are at the nodes of the cavity standing wave, thus where the intracavity intensity vanishes. For $C<0$, instead, the minima are the maxima of the intracavity intensity. 

Finally, for sufficiently deep optical lattices we use the tight-binding and single-band approximation, and obtain a modified Harper's Hamiltonian 
\begin{align}\label{eq:H}
H_\mathrm{TB} =-t \sum_n \left[|n\rangle\langle n+1|+|n \rangle\langle n-1|+\varepsilon_n |n\rangle\langle n|\right]\,,
\end{align}
where $|n \rangle$ is the state vector for the particle localized on the $n$-th site of the lattice, $t$ is hopping integral, and $\varepsilon_n$ scales the on-site energy:
\begin{align}
\label{onsite}
\varepsilon_n= V \mathrm{arctan}(-\delta+C\cos^2(\beta\pi n))\,,
\end{align}
with $V=V_0/t$ the potential depth in units of the tunneling $t$. Hamiltonian \eqref{eq:H} has been obtained in the basis of Wannier functions, which are localized on the optical lattice sites, and discarding long range hopping as well as inhomogeneities in the tunneling coefficients caused by cavity potential $\epsilon(x)$. We remark that, despite the initial model is driven dissipative, in the regime where we can adiabatic eliminate the cavity degrees of freedom the atomic dynamics is strictly Hamiltonian, and thus different from the model considered in Ref. \cite{Zheng18}. The Hamiltonian description of Eq.~\eqref{eq:H0} is valid as long as the coarse-graining discussed above and in Refs. \cite{Larson08,Fernandez10,Habibian13,Schuetz13} applies.

In the rest of this work we will analyse the spectrum of excitations of Hamiltonian $H_\mathrm{TB}$ as a function of $V$, $C$, and $\delta$. These parameters have specific physical meanings. The dimensionless potential depth $V$ is proportional to the intensity of the pump, and thus to the average number of intracavity photons. The parameters $\delta$ and $|C|$ determine the form of the cavity-induced potential. The cooperativity $C$ determines the strength of the cavity-atom optomechanical coupling. For $|C|\ll 1$ the onsite energy essentially reduces to a single harmonic, however with the new amplitude $V'=|C|V/[2(\delta^2+1)]$ and the shift $-{\rm atan}(\delta)$. The critical value at which localization occurs is found at $V'=2t/\alpha$, 
with $\alpha$ a factor depending on the overlap integral between the Wannier functions and the incommensurate potential in the harmonic limit \cite{Rojan16}. When $|C|\ge 1$, instead, higher harmonics become relevant: the value of $|C|$ and its sign determine the form of the nonlinear potential that the atom experiences. 

We finally remark that the model of Eq. \eqref{eq:H0} is strictly valid when only one atom interacts with the cavity field. Thus, it cannot be extended to a Bose-Einstein condensate of atoms with vanishing scattering length, since the atoms will still experience the cavity-mediated long-range interactions \cite{Larson08,Fernandez10}. Therefore, in this work $C_0=|C|$ is strictly speaking the single-atom cooperativity \cite{Kimble94}. Experiments confining a controllable number of atoms within high-finesse optical resonators using a dipole trap have been recently performed, see for instance Refs. \cite{Miller05,Reimann15,Neuzner16}.


\section{Excitation spectrum and mobility edges\label{sec:Results}}

The ground state properties of the system described by the Hamiltonian \eqref{eq:H} have been analyzed in detail in Ref. \cite{Rojan16}. Due to the fact that the potential contains higher harmonics, the model is not dual. The ground state exhibits nevertheless a transition between extended and localized wave function, whose transition point is shifted with respect to the transition without cavity back-action. In particular, for sufficiently large values of $|C|$, in the localized phase the ground state probability density can exhibit a very small, yet finite, contribution from a constant density offset while the Lyapunov exponent of the exponentially-localized component is a function of the cooperativity. 

\begin{figure}
\begin{center}
\includegraphics[width=8cm]{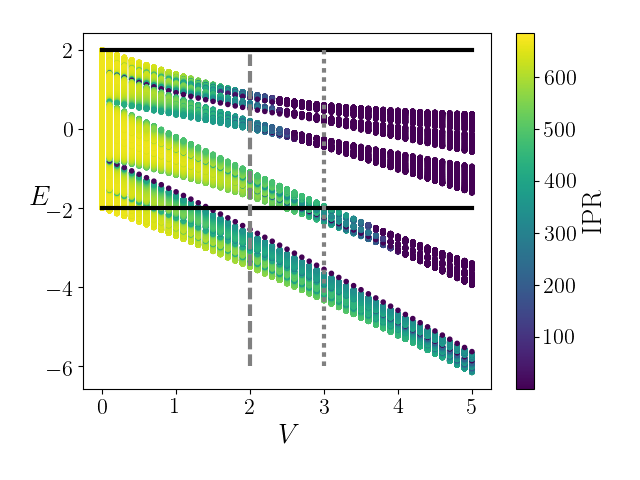}
\caption{(Color online) Eigenenergies (in units of $t$) as a function of the dimensionless potential depth $V=V_0/t$ for $\beta=\phi$, $C=-2$, and $\delta=0$. The dots correspond to the energy values, the color (shade of grey in print) represents their $\IPR$ as visualized by the bar code. The horizontal black lines indicate the bandwidth for $V=0$. The grey vertical lines indicate the parameters used in Fig.~\ref{fig:k}.
\label{fig:cut}}
\end{center}
\end{figure}

In this Section we analyse the properties of the excited states, focussing in particular on identifying a nontrivial mobility edge, namely, an energy eigenstate separating localised and non-localised states within a band \cite{Mott87}. To this end 
we diagonalize the Hamiltonian  \eqref{eq:H}  taking an optical lattice with $N=1000$ sites and open boundary conditions. Using the eigenstates $\psi_j(n)$, where $\psi_j(n)$ is the value of the $j$--eigenstate at the lattice site $n$, we determine the inverse participation ratio $\IPR$ (see e.g. \cite{Thouless74,Kramer93}):
\begin{align}
\label{iprdef}
\mathrm{IPR}_j=\left(\sum_n |\psi_j(n)|^4\right)^{-1}.
\end{align}
From construction $\IPR=1$ for perfect localization, namely, when only one site is occupied. This is the minimal value it can take. The maximal value $\IPR=N$ corresponds to the case of a uniform distribution over the whole lattice. The chosen system size $N=1000$ is sufficiently large to clearly distinguish between extended states, whose $\IPR$ is of order of several hundreds, and localized states. We checked that  states with $\IPR$ values about $10$-$20$ are exponentially localized. Our calculations show that nearly all eigenstates have $\IPR$ which falls into one of these ranges of values. 
An example of the energy spectrum as a function of $V$ is shown in Fig. \ref{fig:cut} for $C=-2$ and $\delta=0$. The dots correspond to the energy values, the color (shade of grey) represents their $\IPR$. One can observe the abrupt change of the $\IPR$ from large to low values (where $\IPR\sim 10$). Moreover, the states of a whole sub-band become localized or extended nearly for the same value of $V$.

The presence of mobility edges can be visualised by introducing the {new} parameter $R$, which is the ratio between the number of localized states (here, states with $\IPR <50$) over $N$ and is defined as
\begin{align}
\label{eq:R}
 R=\frac{\#(\IPR <50)}{ N}\,,
\end{align}
where the threshold $\IPR=50$ has been identified for a lattice of $N=1000$ sites. The ratio $R$ can take all values between 0 and 1, where $R=0$ corresponds to the situation in which all states are extended while for $R=1$ all states are localized. {Thus an abrupt transition between the extreme values of $R$ indicates the localization transition for all the states. Instead, a gradual change of $R$ points towards the existence of the mobility edge where for  given parameter values only part of the eigenstates is localized.}  
\begin{figure}
\begin{center}
\includegraphics[width=8cm]{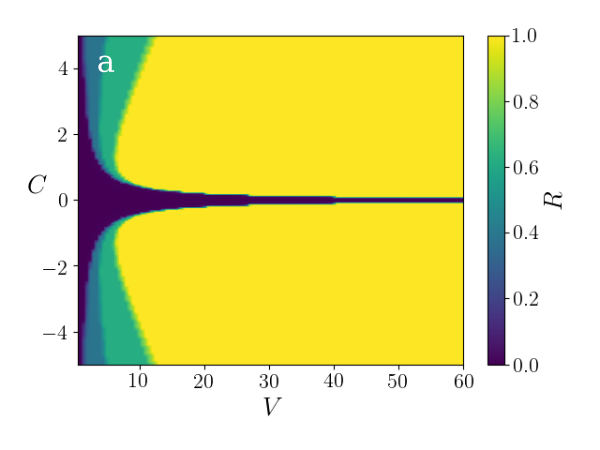}
\includegraphics[width=8cm]{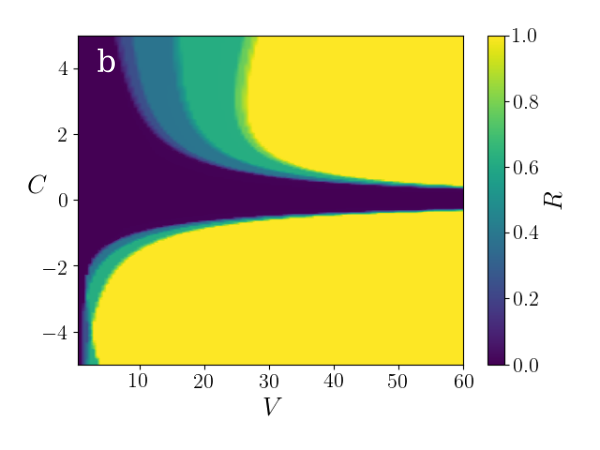}
\caption{(Color online) Contour plot of the ratio of localized states $R$, Eq.~\eqref{eq:R}, as a function of the dimensionless potential depth $V$ and of the cooperativity $C$ for $\beta=\phi=(1+\sqrt{5})/2$  and offsets (a) $\delta=0$ and (b) $\delta=-2$ (see Eq.~\eqref{onsite}).\label{fig:ORG}}
\end{center}
\end{figure}

Figure~\ref{fig:ORG} displays $R$ as a function of the on-site disorder parameters $V$ and $C$ for $\delta=\{0,-2\}$ in Eq.~\ref{onsite}. {For very small $|C|$, {as expected,} all states are extended with vanishing  $R$  - at least at the range of $V$ considered. For stronger  cooperativity $C$, the localization sets in, but the border between $R=0$ and $R=1$ regime is smeared out over significant range of $V$ values. In the region corresponding to intermediate $R$ values only a fraction of states reveals localization. Thus, some of the eigenstates of the system for say, $C\approx4$, $V\approx8$ are localized} {while others are extended. }Since localized and delocalized states cannot coexist at similar energies this behavior indicates the existence of the mobility edge in energies.

{A further insight may be gained comparing Fig.~2(a) with Fig.~2(b) of Ref. \cite{Rojan16}. In the latter the participation ratio, i.e. the inverse of \eqref{iprdef} is plotted (accidentally it is called there as inverse participation ratio in an obvious contradiction with IPR definition \cite{Thouless74,Kramer93}) for the {\it ground} state of the system. As expected the localization border for the ground state is sharp. For positive $C$ it corresponds to the smaller  $V$ border of the transition region  in Fig.~\ref{fig:ORG}(a). Thus for parameters in this region (e.g. $C\approx4$, $V\approx 8$) the ground state is already localized while some excited states are still extended. This proves the existence of the mobility edge in the system, separating the low lying localized states from the higher lying extended states in this transition regime.}

{In this way we can convince ourselves about the existence of a single mobility edge. It may be possible that there exists also an upper mobility edge - not revealed by comparison with the ground state properties. Its detection would require a detailed study of $R$ as a function of the energy, which is beyond the scope of the present paper. Let us note, however, that the mobility edge becomes  "inverted" for negative $C$: Here,  in the parameter regime where $R$ takes already intermediate values,  the ground state may remain extended (as visible by comparing Fig.~\ref{fig:ORG}(a) with  Fig.~2(b) of Ref. \cite{Rojan16} ) thus suggesting localization of some excited states.} 

 For $\delta=-2$, Fig. \ref{fig:ORG}(b), we observe the similar behaviour as for $\delta=0$ with a large transition region  but only for positive $C$ values. For $C<0$ the interval of $V$-values where the mobility edge is found shrinks at $C\sim -4$. We note that for the 
same parameters this is the region where bistability is expected for interacting atoms in the same setup \cite{Fogarty15}. In general, the variation of $R$ with $V$ occurs in steps. This is a consequence of the fact that the energy band splits into flat sub-bands by increasing $V$ from zero, as visible in Fig. \ref{fig:cut}.

\begin{figure}
\begin{center}
\includegraphics[width=8cm]{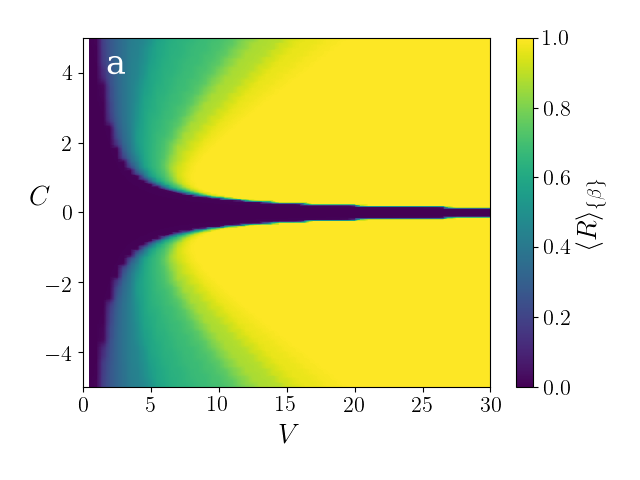}
\includegraphics[width=8cm]{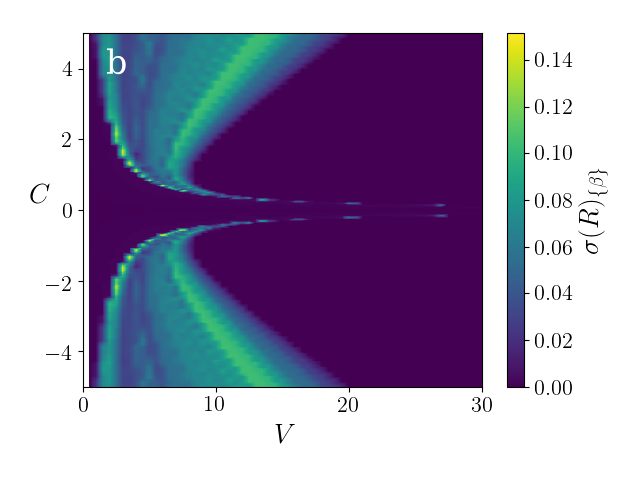}
\caption{(Color online) Contour plots of (a) the average ratio of localized states $\langle R\rangle$, Eq. \eqref{aver} and (b) the standard deviation  $\sigma(R )$ of the distribution, Eq. \eqref{var}, as a function of $C$ and $V$ for offset $\delta=0$ in the potential of Eq.~\eqref{onsite}. The average is taken over $122$ different values of diophantic numbers $\beta=\phi^{ab}_{cd}$, see text for details. \label{fig:D0}}
\end{center}
\end{figure}

\begin{figure}
\begin{center}
\includegraphics[width=8cm]{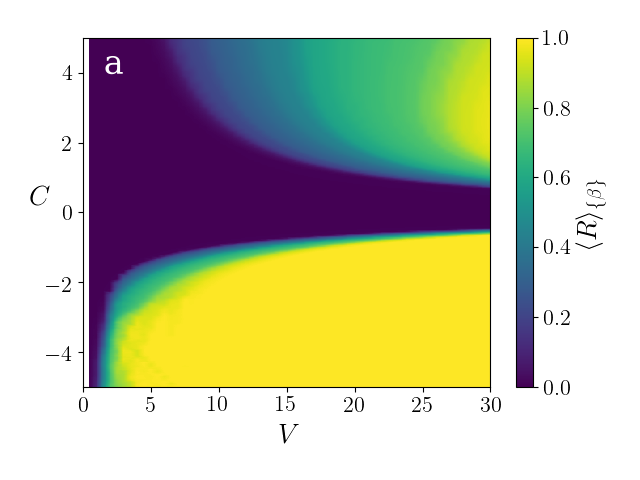}
\includegraphics[width=8cm]{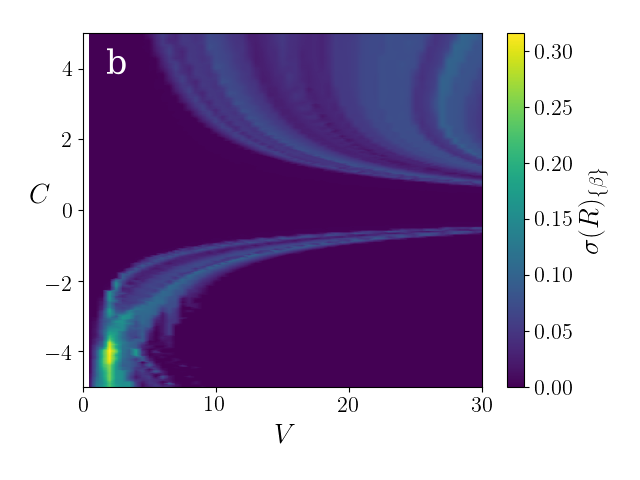}
\caption{(Color online) Same as Fig.~\ref{fig:D0} but for the offset parameter $\delta=-2$ in the potential of Eq.\eqref{onsite} .
\label{fig:D2}}
\end{center}
\end{figure}

So far we discussed the mobility edge for an incommensurability ratio equal to the golden mean, $\beta=\phi=(\sqrt{5}+1)/2$, which is traditionally used in most of the studies in the field. The proof of the localization in the Aubry-Andr\'e model given in Ref. \cite {Jitomirskaya99} is derived for any diophantic number, the number ``as much incommensurate'' as golden ratio -- having the same expansion into continuous fraction from some point. We will now verify whether any diophantic number gives the same results for the potential we are considering, Eq. \eqref{onsite}. For this purpose we use the formula
\begin{align}
\phi^{ab}_{cd}=\frac{a+b\phi}{c+d\phi}\,,
\end{align}
which delivers a whole  (infinite, countable) family of diophantic numbers when $a$, $b$, $c$ and $d$ are integers fulfilling relation $ad-bc=\pm1$ and $\phi=(1+\sqrt{5})/2$ is the golden ratio \cite{Hurwitz91}.
We construct a set of diophantic numbers $\phi^{ab}_{cd}$ in the following way. We take $\ell\in\{2,\ldots,15\}$. For each $\ell$ we find the set of all divisors of $\ell$, $D={D_1,\ldots}$, and of $\ell-1$, $D'={D'_1,\ldots}$. Then we construct the subsequent $\phi^{ab}_{cd}$-s by taking $a=D_i$, $b=D'_j$, $c=(\ell-1)/D'_j$, and $d=\ell/D_i$, and by eliminating reappearing configurations. This yields a set 
of $M=122$ diophantic  numbers $\phi^{ab}_{cd}$. We then evaluate $R(\beta)$ for each value of $\beta=\phi^{ab}_{cd}$ from this set and determine the average $\langle R\rangle_{\{\beta\}}$ and its standard deviation $\sigma(R)_{\{\beta\}}$,  defined as 
\begin{align}\label{aver}
\langle R\rangle_{\{\beta\}} = \frac{1}{M} \sum_\beta R(\beta)\,,\\
\sigma(R) _{\{\beta\}}= \sqrt{\langle R^2\rangle - {\langle R \rangle}^2}\,.
\label{var}
\end{align}
The mean $\langle R\rangle_{\{\beta\}} $ is shown in top panels of Fig.~\ref{fig:D0} and Fig.~\ref{fig:D2}  as a function of $V$ and $C$ fo two different $\delta$ values. 
{A comparison of the mean plots with Fig.~\ref{fig:ORG}  shows that the parameter regions 
corresponding to all the states being extended or being localized is not sensitive to  changes of the incommensurate ratio (note that the horizontal scale is smaller in Fig.~\ref{fig:D0} and Fig.~\ref{fig:D2} than in Fig.~\ref{fig:ORG} for better visibility of the details of the transition region). On the other hand, the contours of the intermediate regime, where the mobility edge appears, are smoothened.}
This indicates that the position of the mobility edge depends  on the specific incommensurability ratio. To our knowledge, this feature has not been reported in other extensions of the Aubry-Andr\'e model before.

{To see whether this effect is really important let us consider the standard deviation 
$\sigma(R) _{\{\beta\}}$ of $R$ distribution shown in lower panels in  Fig.~\ref{fig:D0} and Fig.~\ref{fig:D2}. Clearly deep in the localized regime (where $R\approx 1$) and in the extended regime
(where $R\approx 0$) also the standard deviation takes very small values. By comparison
$\sigma(R) _{\{\beta\}}$ is significant in the transition regime between extended and localized states.
Thus indeed in this regime the behavior of the system is sensitive to the incommensurability ratio.}

\begin{figure}
\begin{center}
\includegraphics[width=8cm]{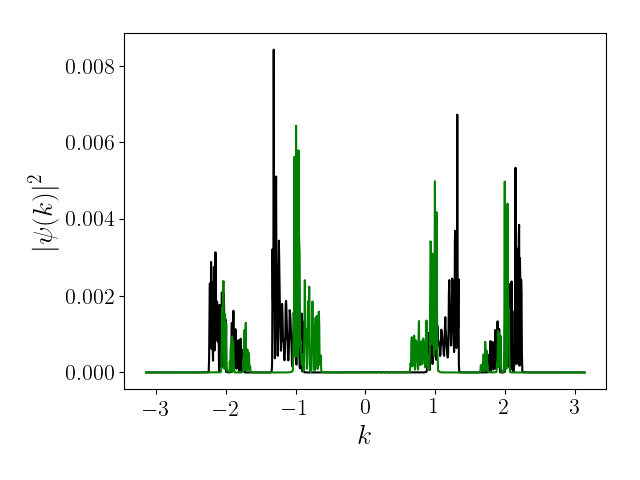}
\caption{(Color online) Distribution $|\psi(k)|^2$ of momenta of waves, leaving the central quasi-disordered region of the system, as a function of $k$. The distribution is evaluated numerically after evolving an initially localized wave function for a time $T=400/t$. The parameters are $\delta=0$, $C=-2$ while $V=2$ and $V=3$ for black and green lines, respectively. The atom is initially at the center of the quasidisordered potential. See text for further details. \label{fig:k}}
\end{center}
\end{figure}

{Let us mention, finally, that t}he existence of controlled mobility edges could be utilized for transport-like experiment, in the spirit of the proposal in Ref. \cite{Plodzien11} where Anderson localization in a  spatially  correlated  disorder potential is shown to operate as a bandpass filter, which selects atoms with certain momenta. In order to show these dynamics in our system, we simulate the time evolution in our model assuming that the atom initially occupies a single site at the center of the system. We assume a quasiperiodic lattice of 100 sites, surrounded by two ``empty'' regions on both sides,  both consisting of 500 sites. The size of the central region is chosen to exceed the localization length. The numerical evolution is calculated using a fourth-order Adams predictor-corrector method up to  time $T=400/t$: This time duration is sufficiently long to allow for the extended component of the initial wave-function to leave the central region. In order to obtain the distribution of momenta $\psi(k)$ of waves that managed to 
escape from the central region, we set to zero the part of the wave-function localized in the central region and take the Fourier transform of the remaining part. Figure~\ref{fig:k} displays $|\psi(k)|^2$ as a function of $k$ at time $T$.
It can be seen that manipulating the system parameters one can select waves {within a certain window of momenta} making it possible to use this class of systems as filters for momenta of particles \cite{Plodzien11}. We do not have analytical description in this case (as possible for some classes of disordered potentials \cite{Major16}). On the other hand  the advantage of the system presented here lies in the fact that the disorder is quasi-random and, therefore, may be reproducible.

\section{Conclusions\label{sec:Conclusions}}

Starting from the model proposed in \cite{Rojan16} we have given a description of the localization properties by determining the full spectrum as a function of the system parameters. We have shown that there exists a finite range of parameters where a mobility edge exists -- some states are localized while other remain extended. We have also shown that fully localized or fully extended states exist in regions of the parameters space, and that these regions are quite robust upon changing the incommensurability ratio $\beta$. Yet, the region where the mobility edge exists is strongly influenced by the value of $\beta$, even if this is any other diophantic number than the golden ratio. In particular, both the position of the mobility edge in parameter space, as well as the fraction of localized states, strongly depend on the chosen incommensurability ratio $\beta$ as revealed by a significant variance of this fraction when different incommensurate parameters are taken. Finally we have discussed a possible 
application of the sensitivity on $\beta$ as a filter for momenta of ultracold atoms. 

We note that the localization 
properties physically originate from the back-action of the cavity field on the atom. The field leaking at the cavity mirrors, moreover, contain information about the atomic state and dynamical properties, this property has been successfully applied for instance for measuring Bloch oscillations \cite{Venkatesh09,Venkatesh13,Georges17}. Future work will focus on the characterization of coherence properties of the emitted light in order to identify and monitor the localization properties.
After finishing this work we became aware of a recent preprint \cite{Li18} discussing mobility edges for other types of incommensurate lattice potentials.

\section*{Acknowlegments}
The authors thank T. Fogarty, A. Minguzzi, K. Rojan, and especially Hessam Habibian and Rebecca Kraus, for discussions. J. M. and J. Z. acknowledge support by  National Science Centre, Poland via projects   OPUS: 2016/21/B/ST2/01086 and  QTFLAG: 2017/25/Z/ST2/03029. Support by PL-Grid Infrastructure, by EU via project QUIC (H2020-FETPROACT-2014 No. 641122), by the DFG DACH project ("Quantum crystals of matter and light"), by the Quantera network "NAQUAS"  is  also acknowledged. Project NAQUAS has received funding from the QuantERA ERA-NET Cofund in Quantum Technologies implemented within the European Union's Horizon 2020 Programme.

\begin{thebibliography}{90}%
\makeatletter
\providecommand \@ifxundefined [1]{%
 \@ifx{#1\undefined}
}%
\providecommand \@ifnum [1]{%
 \ifnum #1\expandafter \@firstoftwo
 \else \expandafter \@secondoftwo
 \fi
}%
\providecommand \@ifx [1]{%
 \ifx #1\expandafter \@firstoftwo
 \else \expandafter \@secondoftwo
 \fi
}%
\providecommand \natexlab [1]{#1}%
\providecommand \enquote  [1]{``#1''}%
\providecommand \bibnamefont  [1]{#1}%
\providecommand \bibfnamefont [1]{#1}%
\providecommand \citenamefont [1]{#1}%
\providecommand \href@noop [0]{\@secondoftwo}%
\providecommand \href [0]{\begingroup \@sanitize@url \@href}%
\providecommand \@href[1]{\@@startlink{#1}\@@href}%
\providecommand \@@href[1]{\endgroup#1\@@endlink}%
\providecommand \@sanitize@url [0]{\catcode `\\12\catcode `\$12\catcode
  `\&12\catcode `\#12\catcode `\^12\catcode `\_12\catcode `\%12\relax}%
\providecommand \@@startlink[1]{}%
\providecommand \@@endlink[0]{}%
\providecommand \url  [0]{\begingroup\@sanitize@url \@url }%
\providecommand \@url [1]{\endgroup\@href {#1}{\urlprefix }}%
\providecommand \urlprefix  [0]{URL }%
\providecommand \Eprint [0]{\href }%
\providecommand \doibase [0]{http://dx.doi.org/}%
\providecommand \selectlanguage [0]{\@gobble}%
\providecommand \bibinfo  [0]{\@secondoftwo}%
\providecommand \bibfield  [0]{\@secondoftwo}%
\providecommand \translation [1]{[#1]}%
\providecommand \BibitemOpen [0]{}%
\providecommand \bibitemStop [0]{}%
\providecommand \bibitemNoStop [0]{.\EOS\space}%
\providecommand \EOS [0]{\spacefactor3000\relax}%
\providecommand \BibitemShut  [1]{\csname bibitem#1\endcsname}%
\let\auto@bib@innerbib\@empty
\bibitem [{\citenamefont {Aubry}\ and\ \citenamefont
  {Andr\'e}(1980)}]{Aubry80}%
  \BibitemOpen
  \bibfield  {author} {\bibinfo {author} {\bibfnamefont {S.}~\bibnamefont
  {Aubry}}\ and\ \bibinfo {author} {\bibfnamefont {G.}~\bibnamefont
  {Andr\'e}},\ }\href@noop {} {\bibfield  {journal} {\bibinfo  {journal} {Ann.
  Israel Phys. Soc.}\ }\textbf {\bibinfo {volume} {3}},\ \bibinfo {pages} {133}
  (\bibinfo {year} {1980})}\BibitemShut {NoStop}%
\bibitem [{\citenamefont {Anderson}(1958)}]{Anderson58}%
  \BibitemOpen
  \bibfield  {author} {\bibinfo {author} {\bibfnamefont {P.}~\bibnamefont
  {Anderson}},\ }\href@noop {} {\bibfield  {journal} {\bibinfo  {journal}
  {Phys. Rev.}\ }\textbf {\bibinfo {volume} {109}},\ \bibinfo {pages} {1492}
  (\bibinfo {year} {1958})}\BibitemShut {NoStop}%
\bibitem [{\citenamefont {Jitomirskaya}(1999)}]{Jitomirskaya99}%
  \BibitemOpen
  \bibfield  {author} {\bibinfo {author} {\bibfnamefont {S.~Y.}\ \bibnamefont
  {Jitomirskaya}},\ }\href {http://www.jstor.org/stable/121066} {\bibfield
  {journal} {\bibinfo  {journal} {Ann. Math.}\ }\textbf {\bibinfo {volume}
  {150}},\ \bibinfo {pages} {1159} (\bibinfo {year} {1999})}\BibitemShut
  {NoStop}%
\bibitem [{\citenamefont {Bloch}\ \emph {et~al.}(2008)\citenamefont {Bloch},
  \citenamefont {Dalibard},\ and\ \citenamefont {Zwerger}}]{Bloch08}%
  \BibitemOpen
  \bibfield  {author} {\bibinfo {author} {\bibfnamefont {I.}~\bibnamefont
  {Bloch}}, \bibinfo {author} {\bibfnamefont {J.}~\bibnamefont {Dalibard}}, \
  and\ \bibinfo {author} {\bibfnamefont {W.}~\bibnamefont {Zwerger}},\ }\href
  {\doibase 10.1103/RevModPhys.80.885} {\bibfield  {journal} {\bibinfo
  {journal} {Rev. Mod. Phys.}\ }\textbf {\bibinfo {volume} {80}},\ \bibinfo
  {pages} {885} (\bibinfo {year} {2008})}\BibitemShut {NoStop}%
\bibitem [{\citenamefont {Jaksch}\ and\ \citenamefont
  {Zoller}(2005)}]{Jaksch2005}%
  \BibitemOpen
  \bibfield  {author} {\bibinfo {author} {\bibfnamefont {D.}~\bibnamefont
  {Jaksch}}\ and\ \bibinfo {author} {\bibfnamefont {P.}~\bibnamefont
  {Zoller}},\ }\href@noop {} {\bibfield  {journal} {\bibinfo  {journal} {Ann.
  Phys.}\ }\textbf {\bibinfo {volume} {315}},\ \bibinfo {pages} {52} (\bibinfo
  {year} {2005})}\BibitemShut {NoStop}%
\bibitem [{\citenamefont {Lewenstein}\ \emph {et~al.}(2012)\citenamefont
  {Lewenstein}, \citenamefont {Sanpera},\ and\ \citenamefont
  {Ahufinger}}]{Lewenstein12}%
  \BibitemOpen
  \bibfield  {author} {\bibinfo {author} {\bibfnamefont {M.}~\bibnamefont
  {Lewenstein}}, \bibinfo {author} {\bibfnamefont {A.}~\bibnamefont {Sanpera}},
  \ and\ \bibinfo {author} {\bibfnamefont {V.}~\bibnamefont {Ahufinger}},\
  }\href@noop {} {\emph {\bibinfo {title} {Ultracold Atoms in Optical Lattices:
  Simulating Many-Body Quantum Systems}}}\ (\bibinfo  {publisher} {Oxford
  University Press},\ \bibinfo {year} {2012})\BibitemShut {NoStop}%
\bibitem [{\citenamefont {Windpassinger}\ and\ \citenamefont
  {Sengstock}(2013)}]{Windpassinger13}%
  \BibitemOpen
  \bibfield  {author} {\bibinfo {author} {\bibfnamefont {P.}~\bibnamefont
  {Windpassinger}}\ and\ \bibinfo {author} {\bibfnamefont {K.}~\bibnamefont
  {Sengstock}},\ }\href {\doibase 10.1088/0034-4885/76/8/086401} {\bibfield
  {journal} {\bibinfo  {journal} {Rep. Prog. Phys.}\ }\textbf {\bibinfo
  {volume} {76}},\ \bibinfo {pages} {086401} (\bibinfo {year}
  {2013})}\BibitemShut {NoStop}%
\bibitem [{\citenamefont {Eckardt}\ \emph {et~al.}(2005)\citenamefont
  {Eckardt}, \citenamefont {Weiss},\ and\ \citenamefont
  {Holthaus}}]{holthaus2005}%
  \BibitemOpen
  \bibfield  {author} {\bibinfo {author} {\bibfnamefont {A.}~\bibnamefont
  {Eckardt}}, \bibinfo {author} {\bibfnamefont {C.}~\bibnamefont {Weiss}}, \
  and\ \bibinfo {author} {\bibfnamefont {M.}~\bibnamefont {Holthaus}},\ }\href
  {\doibase 10.1103/PhysRevLett.95.260404} {\bibfield  {journal} {\bibinfo
  {journal} {Phys. Rev. Lett.}\ }\textbf {\bibinfo {volume} {95}},\ \bibinfo
  {pages} {260404} (\bibinfo {year} {2005})}\BibitemShut {NoStop}%
\bibitem [{\citenamefont {Lignier}\ \emph {et~al.}(2007)\citenamefont
  {Lignier}, \citenamefont {Sias}, \citenamefont {Ciampini}, \citenamefont
  {Singh}, \citenamefont {Zenesini}, \citenamefont {Morsch},\ and\
  \citenamefont {Arimondo}}]{Lignier2007}%
  \BibitemOpen
  \bibfield  {author} {\bibinfo {author} {\bibfnamefont {H.}~\bibnamefont
  {Lignier}}, \bibinfo {author} {\bibfnamefont {C.}~\bibnamefont {Sias}},
  \bibinfo {author} {\bibfnamefont {D.}~\bibnamefont {Ciampini}}, \bibinfo
  {author} {\bibfnamefont {Y.}~\bibnamefont {Singh}}, \bibinfo {author}
  {\bibfnamefont {A.}~\bibnamefont {Zenesini}}, \bibinfo {author}
  {\bibfnamefont {O.}~\bibnamefont {Morsch}}, \ and\ \bibinfo {author}
  {\bibfnamefont {E.}~\bibnamefont {Arimondo}},\ }\href {\doibase
  10.1103/PhysRevLett.99.220403} {\bibfield  {journal} {\bibinfo  {journal}
  {Phys. Rev. Lett.}\ }\textbf {\bibinfo {volume} {99}},\ \bibinfo {pages}
  {220403} (\bibinfo {year} {2007})}\BibitemShut {NoStop}%
\bibitem [{\citenamefont {Struck}\ \emph {et~al.}(2012)\citenamefont {Struck},
  \citenamefont {\"Olschl\"ager}, \citenamefont {Weinberg}, \citenamefont
  {Hauke}, \citenamefont {Simonet}, \citenamefont {Eckardt}, \citenamefont
  {Lewenstein}, \citenamefont {Sengstock},\ and\ \citenamefont
  {Windpassinger}}]{Sengstock12}%
  \BibitemOpen
  \bibfield  {author} {\bibinfo {author} {\bibfnamefont {J.}~\bibnamefont
  {Struck}}, \bibinfo {author} {\bibfnamefont {C.}~\bibnamefont
  {\"Olschl\"ager}}, \bibinfo {author} {\bibfnamefont {M.}~\bibnamefont
  {Weinberg}}, \bibinfo {author} {\bibfnamefont {P.}~\bibnamefont {Hauke}},
  \bibinfo {author} {\bibfnamefont {J.}~\bibnamefont {Simonet}}, \bibinfo
  {author} {\bibfnamefont {A.}~\bibnamefont {Eckardt}}, \bibinfo {author}
  {\bibfnamefont {M.}~\bibnamefont {Lewenstein}}, \bibinfo {author}
  {\bibfnamefont {K.}~\bibnamefont {Sengstock}}, \ and\ \bibinfo {author}
  {\bibfnamefont {P.}~\bibnamefont {Windpassinger}},\ }\href {\doibase
  10.1103/PhysRevLett.108.225304} {\bibfield  {journal} {\bibinfo  {journal}
  {Phys. Rev. Lett.}\ }\textbf {\bibinfo {volume} {108}},\ \bibinfo {pages}
  {225304} (\bibinfo {year} {2012})}\BibitemShut {NoStop}%
\bibitem [{\citenamefont {Sacha}\ \emph {et~al.}(2012)\citenamefont {Sacha},
  \citenamefont {Targo\ifmmode~\acute{n}\else \'{n}\fi{}ska},\ and\
  \citenamefont {Zakrzewski}}]{Targonska12}%
  \BibitemOpen
  \bibfield  {author} {\bibinfo {author} {\bibfnamefont {K.}~\bibnamefont
  {Sacha}}, \bibinfo {author} {\bibfnamefont {K.}~\bibnamefont
  {Targo\ifmmode~\acute{n}\else \'{n}\fi{}ska}}, \ and\ \bibinfo {author}
  {\bibfnamefont {J.}~\bibnamefont {Zakrzewski}},\ }\href {\doibase
  10.1103/PhysRevA.85.053613} {\bibfield  {journal} {\bibinfo  {journal} {Phys.
  Rev. A}\ }\textbf {\bibinfo {volume} {85}},\ \bibinfo {pages} {053613}
  (\bibinfo {year} {2012})}\BibitemShut {NoStop}%
\bibitem [{\citenamefont {Eckardt}\ \emph {et~al.}(2010)\citenamefont
  {Eckardt}, \citenamefont {Hauke}, \citenamefont {Soltan-Panahi},
  \citenamefont {Becker}, \citenamefont {Sengstock},\ and\ \citenamefont
  {Lewenstein}}]{Eckardt10}%
  \BibitemOpen
  \bibfield  {author} {\bibinfo {author} {\bibfnamefont {A.}~\bibnamefont
  {Eckardt}}, \bibinfo {author} {\bibfnamefont {P.}~\bibnamefont {Hauke}},
  \bibinfo {author} {\bibfnamefont {P.}~\bibnamefont {Soltan-Panahi}}, \bibinfo
  {author} {\bibfnamefont {C.}~\bibnamefont {Becker}}, \bibinfo {author}
  {\bibfnamefont {K.}~\bibnamefont {Sengstock}}, \ and\ \bibinfo {author}
  {\bibfnamefont {M.}~\bibnamefont {Lewenstein}},\ }\href {\doibase
  10.1209/0295-5075/89/10010} {\bibfield  {journal} {\bibinfo  {journal} {EPL}\
  }\textbf {\bibinfo {volume} {89}},\ \bibinfo {pages} {10010} (\bibinfo {year}
  {2010})}\BibitemShut {NoStop}%
\bibitem [{\citenamefont {Rapp}\ \emph {et~al.}(2012)\citenamefont {Rapp},
  \citenamefont {Deng},\ and\ \citenamefont {Santos}}]{Rapp12}%
  \BibitemOpen
  \bibfield  {author} {\bibinfo {author} {\bibfnamefont {A.}~\bibnamefont
  {Rapp}}, \bibinfo {author} {\bibfnamefont {X.}~\bibnamefont {Deng}}, \ and\
  \bibinfo {author} {\bibfnamefont {L.}~\bibnamefont {Santos}},\ }\href
  {\doibase 10.1103/PhysRevLett.109.203005} {\bibfield  {journal} {\bibinfo
  {journal} {Phys. Rev. Lett.}\ }\textbf {\bibinfo {volume} {109}},\ \bibinfo
  {pages} {203005} (\bibinfo {year} {2012})}\BibitemShut {NoStop}%
\bibitem [{\citenamefont {Przysi\k{e}\.zna}\ \emph {et~al.}(2015)\citenamefont
  {Przysi\k{e}\.zna}, \citenamefont {Dutta},\ and\ \citenamefont
  {Zakrzewski}}]{przysiezna2015}%
  \BibitemOpen
  \bibfield  {author} {\bibinfo {author} {\bibfnamefont {A.}~\bibnamefont
  {Przysi\k{e}\.zna}}, \bibinfo {author} {\bibfnamefont {O.}~\bibnamefont
  {Dutta}}, \ and\ \bibinfo {author} {\bibfnamefont {J.}~\bibnamefont
  {Zakrzewski}},\ }\href {http://stacks.iop.org/1367-2630/17/i=1/a=013018}
  {\bibfield  {journal} {\bibinfo  {journal} {New J. Phys.}\ }\textbf {\bibinfo
  {volume} {17}},\ \bibinfo {pages} {013018} (\bibinfo {year}
  {2015})}\BibitemShut {NoStop}%
\bibitem [{\citenamefont {{Dutta, O. and Gajda, M. and Hauke, P. and
  Lewenstein, M. and L\"uhmann, D.-S. and Malomed, B. A. and Sowi\'nski, T. and
  Zakrzewski, J.}}(2015)}]{Dutta2015}%
  \BibitemOpen
  \bibfield  {author} {\bibinfo {author} {\bibnamefont {{Dutta, O. and Gajda,
  M. and Hauke, P. and Lewenstein, M. and L\"uhmann, D.-S. and Malomed, B. A.
  and Sowi\'nski, T. and Zakrzewski, J.}}},\ }\href
  {http://stacks.iop.org/0034-4885/78/i=6/a=066001} {\bibfield  {journal}
  {\bibinfo  {journal} {Rep. Prog. Phys.}\ }\textbf {\bibinfo {volume} {78}},\
  \bibinfo {pages} {066001} (\bibinfo {year} {2015})}\BibitemShut {NoStop}%
\bibitem [{\citenamefont {Major}\ \emph {et~al.}(2017)\citenamefont {Major},
  \citenamefont {P\l{}odzie\ifmmode~\acute{n}\else \'{n}\fi{}}, \citenamefont
  {Dutta},\ and\ \citenamefont {Zakrzewski}}]{Major17}%
  \BibitemOpen
  \bibfield  {author} {\bibinfo {author} {\bibfnamefont {J.}~\bibnamefont
  {Major}}, \bibinfo {author} {\bibfnamefont {M.}~\bibnamefont
  {P\l{}odzie\ifmmode~\acute{n}\else \'{n}\fi{}}}, \bibinfo {author}
  {\bibfnamefont {O.}~\bibnamefont {Dutta}}, \ and\ \bibinfo {author}
  {\bibfnamefont {J.}~\bibnamefont {Zakrzewski}},\ }\href {\doibase
  10.1103/PhysRevA.96.033620} {\bibfield  {journal} {\bibinfo  {journal} {Phys.
  Rev. A}\ }\textbf {\bibinfo {volume} {96}},\ \bibinfo {pages} {033620}
  (\bibinfo {year} {2017})}\BibitemShut {NoStop}%
\bibitem [{\citenamefont {Dutta}\ \emph {et~al.}(2017)\citenamefont {Dutta},
  \citenamefont {Tagliacozzo}, \citenamefont {Lewenstein},\ and\ \citenamefont
  {Zakrzewski}}]{Dutta17}%
  \BibitemOpen
  \bibfield  {author} {\bibinfo {author} {\bibfnamefont {O.}~\bibnamefont
  {Dutta}}, \bibinfo {author} {\bibfnamefont {L.}~\bibnamefont {Tagliacozzo}},
  \bibinfo {author} {\bibfnamefont {M.}~\bibnamefont {Lewenstein}}, \ and\
  \bibinfo {author} {\bibfnamefont {J.}~\bibnamefont {Zakrzewski}},\ }\href
  {\doibase 10.1103/PhysRevA.95.053608} {\bibfield  {journal} {\bibinfo
  {journal} {Phys. Rev. A}\ }\textbf {\bibinfo {volume} {95}},\ \bibinfo
  {pages} {053608} (\bibinfo {year} {2017})}\BibitemShut {NoStop}%
\bibitem [{\citenamefont {Chin}\ \emph {et~al.}(2010)\citenamefont {Chin},
  \citenamefont {Grimm}, \citenamefont {Julienne},\ and\ \citenamefont
  {Tiesinga}}]{Chin10}%
  \BibitemOpen
  \bibfield  {author} {\bibinfo {author} {\bibfnamefont {C.}~\bibnamefont
  {Chin}}, \bibinfo {author} {\bibfnamefont {R.}~\bibnamefont {Grimm}},
  \bibinfo {author} {\bibfnamefont {P.}~\bibnamefont {Julienne}}, \ and\
  \bibinfo {author} {\bibfnamefont {E.}~\bibnamefont {Tiesinga}},\ }\href
  {\doibase 10.1103/RevModPhys.82.1225} {\bibfield  {journal} {\bibinfo
  {journal} {Rev. Mod. Phys.}\ }\textbf {\bibinfo {volume} {82}},\ \bibinfo
  {pages} {1225} (\bibinfo {year} {2010})}\BibitemShut {NoStop}%
\bibitem [{\citenamefont {Damski}\ \emph {et~al.}(2003)\citenamefont {Damski},
  \citenamefont {Zakrzewski}, \citenamefont {Santos}, \citenamefont {Zoller},\
  and\ \citenamefont {Lewenstein}}]{Damski03}%
  \BibitemOpen
  \bibfield  {author} {\bibinfo {author} {\bibfnamefont {B.}~\bibnamefont
  {Damski}}, \bibinfo {author} {\bibfnamefont {J.}~\bibnamefont {Zakrzewski}},
  \bibinfo {author} {\bibfnamefont {L.}~\bibnamefont {Santos}}, \bibinfo
  {author} {\bibfnamefont {P.}~\bibnamefont {Zoller}}, \ and\ \bibinfo {author}
  {\bibfnamefont {M.}~\bibnamefont {Lewenstein}},\ }\href@noop {} {\bibfield
  {journal} {\bibinfo  {journal} {Phys. Rev. Lett.}\ }\textbf {\bibinfo
  {volume} {91}},\ \bibinfo {pages} {080403} (\bibinfo {year}
  {2003})}\BibitemShut {NoStop}%
\bibitem [{\citenamefont {Roth}\ and\ \citenamefont {Burnett}(2003)}]{Roth03}%
  \BibitemOpen
  \bibfield  {author} {\bibinfo {author} {\bibfnamefont {R.}~\bibnamefont
  {Roth}}\ and\ \bibinfo {author} {\bibfnamefont {K.}~\bibnamefont {Burnett}},\
  }\href {http://stacks.iop.org/1464-4266/5/i=2/a=358} {\bibfield  {journal}
  {\bibinfo  {journal} {Journal of Optics B: Quantum and Semiclassical Optics}\
  }\textbf {\bibinfo {volume} {5}},\ \bibinfo {pages} {S50} (\bibinfo {year}
  {2003})}\BibitemShut {NoStop}%
\bibitem [{\citenamefont {Lye}\ \emph {et~al.}(2005)\citenamefont {Lye},
  \citenamefont {Fallani}, \citenamefont {Modugno}, \citenamefont {Wiersma},
  \citenamefont {Fort},\ and\ \citenamefont {Inguscio}}]{Lye05}%
  \BibitemOpen
  \bibfield  {author} {\bibinfo {author} {\bibfnamefont {J.~E.}\ \bibnamefont
  {Lye}}, \bibinfo {author} {\bibfnamefont {L.}~\bibnamefont {Fallani}},
  \bibinfo {author} {\bibfnamefont {M.}~\bibnamefont {Modugno}}, \bibinfo
  {author} {\bibfnamefont {D.~S.}\ \bibnamefont {Wiersma}}, \bibinfo {author}
  {\bibfnamefont {C.}~\bibnamefont {Fort}}, \ and\ \bibinfo {author}
  {\bibfnamefont {M.}~\bibnamefont {Inguscio}},\ }\href {\doibase
  10.1103/PhysRevLett.95.070401} {\bibfield  {journal} {\bibinfo  {journal}
  {Phys. Rev. Lett.}\ }\textbf {\bibinfo {volume} {95}},\ \bibinfo {pages}
  {070401} (\bibinfo {year} {2005})}\BibitemShut {NoStop}%
\bibitem [{\citenamefont {Cl\'ement}\ \emph {et~al.}(2005)\citenamefont
  {Cl\'ement}, \citenamefont {Var\'on}, \citenamefont {Hugbart}, \citenamefont
  {Retter}, \citenamefont {Bouyer}, \citenamefont {Sanchez-Palencia},
  \citenamefont {Gangardt}, \citenamefont {Shlyapnikov},\ and\ \citenamefont
  {Aspect}}]{Clement05}%
  \BibitemOpen
  \bibfield  {author} {\bibinfo {author} {\bibfnamefont {D.}~\bibnamefont
  {Cl\'ement}}, \bibinfo {author} {\bibfnamefont {A.~F.}\ \bibnamefont
  {Var\'on}}, \bibinfo {author} {\bibfnamefont {M.}~\bibnamefont {Hugbart}},
  \bibinfo {author} {\bibfnamefont {J.~A.}\ \bibnamefont {Retter}}, \bibinfo
  {author} {\bibfnamefont {P.}~\bibnamefont {Bouyer}}, \bibinfo {author}
  {\bibfnamefont {L.}~\bibnamefont {Sanchez-Palencia}}, \bibinfo {author}
  {\bibfnamefont {D.~M.}\ \bibnamefont {Gangardt}}, \bibinfo {author}
  {\bibfnamefont {G.~V.}\ \bibnamefont {Shlyapnikov}}, \ and\ \bibinfo {author}
  {\bibfnamefont {A.}~\bibnamefont {Aspect}},\ }\href {\doibase
  10.1103/PhysRevLett.95.170409} {\bibfield  {journal} {\bibinfo  {journal}
  {Phys. Rev. Lett.}\ }\textbf {\bibinfo {volume} {95}},\ \bibinfo {pages}
  {170409} (\bibinfo {year} {2005})}\BibitemShut {NoStop}%
\bibitem [{\citenamefont {Fort}\ \emph {et~al.}(2005)\citenamefont {Fort},
  \citenamefont {Fallani}, \citenamefont {Guarrera}, \citenamefont {Lye},
  \citenamefont {Modugno}, \citenamefont {Wiersma},\ and\ \citenamefont
  {Inguscio}}]{Fort05}%
  \BibitemOpen
  \bibfield  {author} {\bibinfo {author} {\bibfnamefont {C.}~\bibnamefont
  {Fort}}, \bibinfo {author} {\bibfnamefont {L.}~\bibnamefont {Fallani}},
  \bibinfo {author} {\bibfnamefont {V.}~\bibnamefont {Guarrera}}, \bibinfo
  {author} {\bibfnamefont {J.~E.}\ \bibnamefont {Lye}}, \bibinfo {author}
  {\bibfnamefont {M.}~\bibnamefont {Modugno}}, \bibinfo {author} {\bibfnamefont
  {D.~S.}\ \bibnamefont {Wiersma}}, \ and\ \bibinfo {author} {\bibfnamefont
  {M.}~\bibnamefont {Inguscio}},\ }\href {\doibase
  10.1103/PhysRevLett.95.170410} {\bibfield  {journal} {\bibinfo  {journal}
  {Phys. Rev. Lett.}\ }\textbf {\bibinfo {volume} {95}},\ \bibinfo {pages}
  {170410} (\bibinfo {year} {2005})}\BibitemShut {NoStop}%
\bibitem [{\citenamefont {Schulte}\ \emph {et~al.}(2005)\citenamefont
  {Schulte}, \citenamefont {Drenkelforth}, \citenamefont {Kruse}, \citenamefont
  {Ertmer}, \citenamefont {Arlt}, \citenamefont {Sacha}, \citenamefont
  {Zakrzewski},\ and\ \citenamefont {Lewenstein}}]{Schulte05}%
  \BibitemOpen
  \bibfield  {author} {\bibinfo {author} {\bibfnamefont {T.}~\bibnamefont
  {Schulte}}, \bibinfo {author} {\bibfnamefont {S.}~\bibnamefont
  {Drenkelforth}}, \bibinfo {author} {\bibfnamefont {J.}~\bibnamefont {Kruse}},
  \bibinfo {author} {\bibfnamefont {W.}~\bibnamefont {Ertmer}}, \bibinfo
  {author} {\bibfnamefont {J.}~\bibnamefont {Arlt}}, \bibinfo {author}
  {\bibfnamefont {K.}~\bibnamefont {Sacha}}, \bibinfo {author} {\bibfnamefont
  {J.}~\bibnamefont {Zakrzewski}}, \ and\ \bibinfo {author} {\bibfnamefont
  {M.}~\bibnamefont {Lewenstein}},\ }\href {\doibase
  10.1103/PhysRevLett.95.170411} {\bibfield  {journal} {\bibinfo  {journal}
  {Phys. Rev. Lett.}\ }\textbf {\bibinfo {volume} {95}},\ \bibinfo {pages}
  {170411} (\bibinfo {year} {2005})}\BibitemShut {NoStop}%
\bibitem [{\citenamefont {Schulte}\ \emph {et~al.}(2006)\citenamefont
  {Schulte}, \citenamefont {Drenkelforth}, \citenamefont {Kruse}, \citenamefont
  {Tiemeyer}, \citenamefont {Sacha}, \citenamefont {Zakrzewski}, \citenamefont
  {Lewenstein}, \citenamefont {Ertmer},\ and\ \citenamefont
  {Arlt}}]{Schulte06}%
  \BibitemOpen
  \bibfield  {author} {\bibinfo {author} {\bibfnamefont {T.}~\bibnamefont
  {Schulte}}, \bibinfo {author} {\bibfnamefont {S.}~\bibnamefont
  {Drenkelforth}}, \bibinfo {author} {\bibfnamefont {J.}~\bibnamefont {Kruse}},
  \bibinfo {author} {\bibfnamefont {R.}~\bibnamefont {Tiemeyer}}, \bibinfo
  {author} {\bibfnamefont {K.}~\bibnamefont {Sacha}}, \bibinfo {author}
  {\bibfnamefont {J.}~\bibnamefont {Zakrzewski}}, \bibinfo {author}
  {\bibfnamefont {M.}~\bibnamefont {Lewenstein}}, \bibinfo {author}
  {\bibfnamefont {W.}~\bibnamefont {Ertmer}}, \ and\ \bibinfo {author}
  {\bibfnamefont {J.~J.}\ \bibnamefont {Arlt}},\ }\href
  {http://stacks.iop.org/1367-2630/8/i=10/a=230} {\bibfield  {journal}
  {\bibinfo  {journal} {New Journal of Physics}\ }\textbf {\bibinfo {volume}
  {8}},\ \bibinfo {pages} {230} (\bibinfo {year} {2006})}\BibitemShut {NoStop}%
\bibitem [{\citenamefont {Billy}\ \emph {et~al.}(2008)\citenamefont {Billy},
  \citenamefont {Josse}, \citenamefont {Zuo}, \citenamefont {Bernard},
  \citenamefont {Hambrecht}, \citenamefont {Lugan}, \citenamefont {Clément},
  \citenamefont {Sanchez-Palencia}, \citenamefont {Bouyer},\ and\ \citenamefont
  {Aspect}}]{Billy08}%
  \BibitemOpen
  \bibfield  {author} {\bibinfo {author} {\bibfnamefont {J.}~\bibnamefont
  {Billy}}, \bibinfo {author} {\bibfnamefont {V.}~\bibnamefont {Josse}},
  \bibinfo {author} {\bibfnamefont {Z.}~\bibnamefont {Zuo}}, \bibinfo {author}
  {\bibfnamefont {A.}~\bibnamefont {Bernard}}, \bibinfo {author} {\bibfnamefont
  {B.}~\bibnamefont {Hambrecht}}, \bibinfo {author} {\bibfnamefont
  {P.}~\bibnamefont {Lugan}}, \bibinfo {author} {\bibfnamefont
  {D.}~\bibnamefont {Clément}}, \bibinfo {author} {\bibfnamefont
  {L.}~\bibnamefont {Sanchez-Palencia}}, \bibinfo {author} {\bibfnamefont
  {P.}~\bibnamefont {Bouyer}}, \ and\ \bibinfo {author} {\bibfnamefont
  {A.}~\bibnamefont {Aspect}},\ }\href@noop {} {\bibfield  {journal} {\bibinfo
  {journal} {Nature}\ }\textbf {\bibinfo {volume} {453}},\ \bibinfo {pages}
  {891} (\bibinfo {year} {2008})}\BibitemShut {NoStop}%
\bibitem [{\citenamefont {Roati}\ \emph {et~al.}(2008)\citenamefont {Roati},
  \citenamefont {D'Errico}, \citenamefont {Fallani}, \citenamefont {Fattori},
  \citenamefont {For}, \citenamefont {Zaccanti}, \citenamefont {Modugno},
  \citenamefont {Modugno},\ and\ \citenamefont {Inguscio}}]{Roati08}%
  \BibitemOpen
  \bibfield  {author} {\bibinfo {author} {\bibfnamefont {G.}~\bibnamefont
  {Roati}}, \bibinfo {author} {\bibfnamefont {C.}~\bibnamefont {D'Errico}},
  \bibinfo {author} {\bibfnamefont {L.}~\bibnamefont {Fallani}}, \bibinfo
  {author} {\bibfnamefont {M.}~\bibnamefont {Fattori}}, \bibinfo {author}
  {\bibfnamefont {C.}~\bibnamefont {For}}, \bibinfo {author} {\bibfnamefont
  {M.}~\bibnamefont {Zaccanti}}, \bibinfo {author} {\bibfnamefont
  {G.}~\bibnamefont {Modugno}}, \bibinfo {author} {\bibfnamefont
  {M.}~\bibnamefont {Modugno}}, \ and\ \bibinfo {author} {\bibfnamefont
  {M.}~\bibnamefont {Inguscio}},\ }\href@noop {} {\bibfield  {journal}
  {\bibinfo  {journal} {Nature}\ }\textbf {\bibinfo {volume} {453}},\ \bibinfo
  {pages} {895} (\bibinfo {year} {2008})}\BibitemShut {NoStop}%
\bibitem [{\citenamefont {Kondov}\ \emph {et~al.}(2011)\citenamefont {Kondov},
  \citenamefont {McGehee}, \citenamefont {Zirbel},\ and\ \citenamefont
  {DeMarco}}]{Kondov11}%
  \BibitemOpen
  \bibfield  {author} {\bibinfo {author} {\bibfnamefont {S.~S.}\ \bibnamefont
  {Kondov}}, \bibinfo {author} {\bibfnamefont {W.~R.}\ \bibnamefont {McGehee}},
  \bibinfo {author} {\bibfnamefont {J.~J.}\ \bibnamefont {Zirbel}}, \ and\
  \bibinfo {author} {\bibfnamefont {B.}~\bibnamefont {DeMarco}},\ }\href@noop
  {} {\bibfield  {journal} {\bibinfo  {journal} {Science}\ }\textbf {\bibinfo
  {volume} {334}},\ \bibinfo {pages} {66} (\bibinfo {year} {2011})}\BibitemShut
  {NoStop}%
\bibitem [{\citenamefont {Jendrzejewski}\ \emph {et~al.}(2012)\citenamefont
  {Jendrzejewski}, \citenamefont {Bernard}, \citenamefont {Mueller},
  \citenamefont {Patrick}, \citenamefont {Josse}, \citenamefont {Piraud},
  \citenamefont {Pezze}, \citenamefont {Sanchez-Palencia}, \citenamefont
  {Aspect},\ and\ \citenamefont {Bouyer}}]{Jendrzejewski12}%
  \BibitemOpen
  \bibfield  {author} {\bibinfo {author} {\bibfnamefont {F.}~\bibnamefont
  {Jendrzejewski}}, \bibinfo {author} {\bibfnamefont {A.}~\bibnamefont
  {Bernard}}, \bibinfo {author} {\bibfnamefont {K.}~\bibnamefont {Mueller}},
  \bibinfo {author} {\bibfnamefont {C.}~\bibnamefont {Patrick}}, \bibinfo
  {author} {\bibfnamefont {V.}~\bibnamefont {Josse}}, \bibinfo {author}
  {\bibfnamefont {M.}~\bibnamefont {Piraud}}, \bibinfo {author} {\bibfnamefont
  {L.}~\bibnamefont {Pezze}}, \bibinfo {author} {\bibfnamefont
  {L.}~\bibnamefont {Sanchez-Palencia}}, \bibinfo {author} {\bibfnamefont
  {A.}~\bibnamefont {Aspect}}, \ and\ \bibinfo {author} {\bibfnamefont
  {P.}~\bibnamefont {Bouyer}},\ }\href@noop {} {\bibfield  {journal} {\bibinfo
  {journal} {Nature Phys.}\ }\textbf {\bibinfo {volume} {8}},\ \bibinfo {pages}
  {398} (\bibinfo {year} {2012})}\BibitemShut {NoStop}%
\bibitem [{\citenamefont {Basko}\ \emph {et~al.}(2006)\citenamefont {Basko},
  \citenamefont {Aleiner},\ and\ \citenamefont {Altschuler}}]{Basko06}%
  \BibitemOpen
  \bibfield  {author} {\bibinfo {author} {\bibfnamefont {D.}~\bibnamefont
  {Basko}}, \bibinfo {author} {\bibfnamefont {I.}~\bibnamefont {Aleiner}}, \
  and\ \bibinfo {author} {\bibfnamefont {B.}~\bibnamefont {Altschuler}},\
  }\href@noop {} {\bibfield  {journal} {\bibinfo  {journal} {Ann. Phys. (NY)}\
  }\textbf {\bibinfo {volume} {321}},\ \bibinfo {pages} {1126} (\bibinfo {year}
  {2006})}\BibitemShut {NoStop}%
\bibitem [{\citenamefont {Oganesyan}\ and\ \citenamefont
  {Huse}(2007)}]{Oganesyan07}%
  \BibitemOpen
  \bibfield  {author} {\bibinfo {author} {\bibfnamefont {V.}~\bibnamefont
  {Oganesyan}}\ and\ \bibinfo {author} {\bibfnamefont {D.~A.}\ \bibnamefont
  {Huse}},\ }\href {\doibase 10.1103/PhysRevB.75.155111} {\bibfield  {journal}
  {\bibinfo  {journal} {Phys. Rev. B}\ }\textbf {\bibinfo {volume} {75}},\
  \bibinfo {pages} {155111} (\bibinfo {year} {2007})}\BibitemShut {NoStop}%
\bibitem [{\citenamefont {Srednicki}(1994)}]{Srednicki94}%
  \BibitemOpen
  \bibfield  {author} {\bibinfo {author} {\bibfnamefont {M.}~\bibnamefont
  {Srednicki}},\ }\href {\doibase 10.1103/PhysRevE.50.888} {\bibfield
  {journal} {\bibinfo  {journal} {Phys. Rev. E}\ }\textbf {\bibinfo {volume}
  {50}},\ \bibinfo {pages} {888} (\bibinfo {year} {1994})}\BibitemShut
  {NoStop}%
\bibitem [{\citenamefont {Huse}\ \emph {et~al.}(2014)\citenamefont {Huse},
  \citenamefont {Nandkishore},\ and\ \citenamefont {Oganesyan}}]{Huse14}%
  \BibitemOpen
  \bibfield  {author} {\bibinfo {author} {\bibfnamefont {D.~A.}\ \bibnamefont
  {Huse}}, \bibinfo {author} {\bibfnamefont {R.}~\bibnamefont {Nandkishore}}, \
  and\ \bibinfo {author} {\bibfnamefont {V.}~\bibnamefont {Oganesyan}},\ }\href
  {\doibase 10.1103/PhysRevB.90.174202} {\bibfield  {journal} {\bibinfo
  {journal} {Phys. Rev. B}\ }\textbf {\bibinfo {volume} {90}},\ \bibinfo
  {pages} {174202} (\bibinfo {year} {2014})}\BibitemShut {NoStop}%
\bibitem [{\citenamefont {Nandkishore}\ and\ \citenamefont
  {Huse}(2015)}]{Rahul15}%
  \BibitemOpen
  \bibfield  {author} {\bibinfo {author} {\bibfnamefont {R.}~\bibnamefont
  {Nandkishore}}\ and\ \bibinfo {author} {\bibfnamefont {D.~A.}\ \bibnamefont
  {Huse}},\ }\href@noop {} {\bibfield  {journal} {\bibinfo  {journal} {Ann.
  Rev. Cond. Mat. Phys.}\ }\textbf {\bibinfo {volume} {6}},\ \bibinfo {pages}
  {15} (\bibinfo {year} {2015})}\BibitemShut {NoStop}%
\bibitem [{\citenamefont {Schreiber}\ \emph {et~al.}(2015)\citenamefont
  {Schreiber}, \citenamefont {Hodgman}, \citenamefont {Bordia}, \citenamefont
  {L\"uschen}, \citenamefont {Fischer}, \citenamefont {Vosk}, \citenamefont
  {Altman}, \citenamefont {Schneider},\ and\ \citenamefont
  {Bloch}}]{Schreiber15}%
  \BibitemOpen
  \bibfield  {author} {\bibinfo {author} {\bibfnamefont {M.}~\bibnamefont
  {Schreiber}}, \bibinfo {author} {\bibfnamefont {S.~S.}\ \bibnamefont
  {Hodgman}}, \bibinfo {author} {\bibfnamefont {P.}~\bibnamefont {Bordia}},
  \bibinfo {author} {\bibfnamefont {H.~P.}\ \bibnamefont {L\"uschen}}, \bibinfo
  {author} {\bibfnamefont {M.~H.}\ \bibnamefont {Fischer}}, \bibinfo {author}
  {\bibfnamefont {R.}~\bibnamefont {Vosk}}, \bibinfo {author} {\bibfnamefont
  {E.}~\bibnamefont {Altman}}, \bibinfo {author} {\bibfnamefont
  {U.}~\bibnamefont {Schneider}}, \ and\ \bibinfo {author} {\bibfnamefont
  {I.}~\bibnamefont {Bloch}},\ }\href@noop {} {\bibfield  {journal} {\bibinfo
  {journal} {Science}\ }\textbf {\bibinfo {volume} {349}},\ \bibinfo {pages}
  {7432} (\bibinfo {year} {2015})}\BibitemShut {NoStop}%
\bibitem [{\citenamefont {{Choi}}\ \emph {et~al.}(2016)\citenamefont {{Choi}},
  \citenamefont {{Hild}}, \citenamefont {{Zeiher}}, \citenamefont
  {{Schau{\ss}}}, \citenamefont {{Rubio-Abadal}}, \citenamefont {{Yefsah}},
  \citenamefont {{Khemani}}, \citenamefont {{Huse}}, \citenamefont {{Bloch}},\
  and\ \citenamefont {{Gross}}}]{Choi16}%
  \BibitemOpen
  \bibfield  {author} {\bibinfo {author} {\bibfnamefont {J.-y.}\ \bibnamefont
  {{Choi}}}, \bibinfo {author} {\bibfnamefont {S.}~\bibnamefont {{Hild}}},
  \bibinfo {author} {\bibfnamefont {J.}~\bibnamefont {{Zeiher}}}, \bibinfo
  {author} {\bibfnamefont {P.}~\bibnamefont {{Schau{\ss}}}}, \bibinfo {author}
  {\bibfnamefont {A.}~\bibnamefont {{Rubio-Abadal}}}, \bibinfo {author}
  {\bibfnamefont {T.}~\bibnamefont {{Yefsah}}}, \bibinfo {author}
  {\bibfnamefont {V.}~\bibnamefont {{Khemani}}}, \bibinfo {author}
  {\bibfnamefont {D.~A.}\ \bibnamefont {{Huse}}}, \bibinfo {author}
  {\bibfnamefont {I.}~\bibnamefont {{Bloch}}}, \ and\ \bibinfo {author}
  {\bibfnamefont {C.}~\bibnamefont {{Gross}}},\ }\href@noop {} {\bibfield
  {journal} {\bibinfo  {journal} {ArXiv e-prints}\ } (\bibinfo {year}
  {2016})},\ \Eprint {http://arxiv.org/abs/1604.04178} {arXiv:1604.04178
  [cond-mat.quant-gas]} \BibitemShut {NoStop}%
\bibitem [{\citenamefont {Sierant}\ \emph {et~al.}(2017)\citenamefont
  {Sierant}, \citenamefont {Delande},\ and\ \citenamefont
  {Zakrzewski}}]{Sierant17}%
  \BibitemOpen
  \bibfield  {author} {\bibinfo {author} {\bibfnamefont {P.}~\bibnamefont
  {Sierant}}, \bibinfo {author} {\bibfnamefont {D.}~\bibnamefont {Delande}}, \
  and\ \bibinfo {author} {\bibfnamefont {J.}~\bibnamefont {Zakrzewski}},\
  }\href {\doibase 10.1103/PhysRevA.95.021601} {\bibfield  {journal} {\bibinfo
  {journal} {Phys. Rev. A}\ }\textbf {\bibinfo {volume} {95}},\ \bibinfo
  {pages} {021601} (\bibinfo {year} {2017})}\BibitemShut {NoStop}%
\bibitem [{\citenamefont {{Sierant}}\ \emph {et~al.}(2017)\citenamefont
  {{Sierant}}, \citenamefont {Delande},\ and\ \citenamefont
  {{Zakrzewski}}}]{Sierant17b}%
  \BibitemOpen
  \bibfield  {author} {\bibinfo {author} {\bibfnamefont {P.}~\bibnamefont
  {{Sierant}}}, \bibinfo {author} {\bibfnamefont {D.}~\bibnamefont {Delande}},
  \ and\ \bibinfo {author} {\bibfnamefont {J.}~\bibnamefont {{Zakrzewski}}},\
  }\href@noop {} {\bibfield  {journal} {\bibinfo  {journal} {Acta Phys. Polon.
  A}\ }\textbf {\bibinfo {volume} {132}},\ \bibinfo {pages} {1707} (\bibinfo
  {year} {2017})}\BibitemShut {NoStop}%
\bibitem [{\citenamefont {Sierant}\ and\ \citenamefont
  {Zakrzewski}(2018)}]{Sierant18}%
  \BibitemOpen
  \bibfield  {author} {\bibinfo {author} {\bibfnamefont {P.}~\bibnamefont
  {Sierant}}\ and\ \bibinfo {author} {\bibfnamefont {J.}~\bibnamefont
  {Zakrzewski}},\ }\href {http://stacks.iop.org/1367-2630/20/i=4/a=043032}
  {\bibfield  {journal} {\bibinfo  {journal} {New Journal of Physics}\ }\textbf
  {\bibinfo {volume} {20}},\ \bibinfo {pages} {043032} (\bibinfo {year}
  {2018})}\BibitemShut {NoStop}%
\bibitem [{\citenamefont {Sacha}\ \emph {et~al.}(2009)\citenamefont {Sacha},
  \citenamefont {M\"uller}, \citenamefont {Delande},\ and\ \citenamefont
  {Zakrzewski}}]{Sacha09}%
  \BibitemOpen
  \bibfield  {author} {\bibinfo {author} {\bibfnamefont {K.}~\bibnamefont
  {Sacha}}, \bibinfo {author} {\bibfnamefont {C.~A.}\ \bibnamefont {M\"uller}},
  \bibinfo {author} {\bibfnamefont {D.}~\bibnamefont {Delande}}, \ and\
  \bibinfo {author} {\bibfnamefont {J.}~\bibnamefont {Zakrzewski}},\ }\href
  {\doibase 10.1103/PhysRevLett.103.210402} {\bibfield  {journal} {\bibinfo
  {journal} {Phys. Rev. Lett.}\ }\textbf {\bibinfo {volume} {103}},\ \bibinfo
  {pages} {210402} (\bibinfo {year} {2009})}\BibitemShut {NoStop}%
\bibitem [{\citenamefont {Evers}\ and\ \citenamefont {Mirlin}(2008)}]{Evers08}%
  \BibitemOpen
  \bibfield  {author} {\bibinfo {author} {\bibfnamefont {F.}~\bibnamefont
  {Evers}}\ and\ \bibinfo {author} {\bibfnamefont {A.~D.}\ \bibnamefont
  {Mirlin}},\ }\href {\doibase 10.1103/RevModPhys.80.1355} {\bibfield
  {journal} {\bibinfo  {journal} {Rev. Mod. Phys.}\ }\textbf {\bibinfo {volume}
  {80}},\ \bibinfo {pages} {1355} (\bibinfo {year} {2008})}\BibitemShut
  {NoStop}%
\bibitem [{\citenamefont {M\"uller}\ \emph {et~al.}(2016)\citenamefont
  {M\"uller}, \citenamefont {Delande},\ and\ \citenamefont
  {Shapiro}}]{Mueller16}%
  \BibitemOpen
  \bibfield  {author} {\bibinfo {author} {\bibfnamefont {C.~A.}\ \bibnamefont
  {M\"uller}}, \bibinfo {author} {\bibfnamefont {D.}~\bibnamefont {Delande}}, \
  and\ \bibinfo {author} {\bibfnamefont {B.}~\bibnamefont {Shapiro}},\ }\href
  {\doibase 10.1103/PhysRevA.94.033615} {\bibfield  {journal} {\bibinfo
  {journal} {Phys. Rev. A}\ }\textbf {\bibinfo {volume} {94}},\ \bibinfo
  {pages} {033615} (\bibinfo {year} {2016})}\BibitemShut {NoStop}%
\bibitem [{\citenamefont {Semeghini}\ \emph {et~al.}(2015)\citenamefont
  {Semeghini}, \citenamefont {Landini}, \citenamefont {Castilho}, \citenamefont
  {Roy}, \citenamefont {Spagnolli}, \citenamefont {Trenkwalder}, \citenamefont
  {Fattori}, \citenamefont {Inguscio},\ and\ \citenamefont
  {Modugno}}]{Semeghini15}%
  \BibitemOpen
  \bibfield  {author} {\bibinfo {author} {\bibfnamefont {G.}~\bibnamefont
  {Semeghini}}, \bibinfo {author} {\bibfnamefont {M.}~\bibnamefont {Landini}},
  \bibinfo {author} {\bibfnamefont {P.}~\bibnamefont {Castilho}}, \bibinfo
  {author} {\bibfnamefont {S.}~\bibnamefont {Roy}}, \bibinfo {author}
  {\bibfnamefont {G.}~\bibnamefont {Spagnolli}}, \bibinfo {author}
  {\bibfnamefont {A.}~\bibnamefont {Trenkwalder}}, \bibinfo {author}
  {\bibfnamefont {M.}~\bibnamefont {Fattori}}, \bibinfo {author} {\bibfnamefont
  {M.}~\bibnamefont {Inguscio}}, \ and\ \bibinfo {author} {\bibfnamefont
  {G.}~\bibnamefont {Modugno}},\ }\href {http://dx.doi.org/10.1038/nphys3339}
  {\bibfield  {journal} {\bibinfo  {journal} {Nature Physics}\ }\textbf
  {\bibinfo {volume} {11}},\ \bibinfo {pages} {554 EP } (\bibinfo {year}
  {2015})},\ \bibinfo {note} {article}\BibitemShut {NoStop}%
\bibitem [{\citenamefont {Pasek}\ \emph {et~al.}(2017)\citenamefont {Pasek},
  \citenamefont {Orso},\ and\ \citenamefont {Delande}}]{Pasek17}%
  \BibitemOpen
  \bibfield  {author} {\bibinfo {author} {\bibfnamefont {M.}~\bibnamefont
  {Pasek}}, \bibinfo {author} {\bibfnamefont {G.}~\bibnamefont {Orso}}, \ and\
  \bibinfo {author} {\bibfnamefont {D.}~\bibnamefont {Delande}},\ }\href
  {\doibase 10.1103/PhysRevLett.118.170403} {\bibfield  {journal} {\bibinfo
  {journal} {Phys. Rev. Lett.}\ }\textbf {\bibinfo {volume} {118}},\ \bibinfo
  {pages} {170403} (\bibinfo {year} {2017})}\BibitemShut {NoStop}%
\bibitem [{\citenamefont {Sacha}\ and\ \citenamefont
  {Delande}(2016)}]{Sacha16}%
  \BibitemOpen
  \bibfield  {author} {\bibinfo {author} {\bibfnamefont {K.}~\bibnamefont
  {Sacha}}\ and\ \bibinfo {author} {\bibfnamefont {D.}~\bibnamefont
  {Delande}},\ }\href {\doibase 10.1103/PhysRevA.94.023633} {\bibfield
  {journal} {\bibinfo  {journal} {Phys. Rev. A}\ }\textbf {\bibinfo {volume}
  {94}},\ \bibinfo {pages} {023633} (\bibinfo {year} {2016})}\BibitemShut
  {NoStop}%
\bibitem [{\citenamefont {Delande}\ \emph {et~al.}(2017)\citenamefont
  {Delande}, \citenamefont {Morales-Molina},\ and\ \citenamefont
  {Sacha}}]{Delande17}%
  \BibitemOpen
  \bibfield  {author} {\bibinfo {author} {\bibfnamefont {D.}~\bibnamefont
  {Delande}}, \bibinfo {author} {\bibfnamefont {L.}~\bibnamefont
  {Morales-Molina}}, \ and\ \bibinfo {author} {\bibfnamefont {K.}~\bibnamefont
  {Sacha}},\ }\href {\doibase 10.1103/PhysRevLett.119.230404} {\bibfield
  {journal} {\bibinfo  {journal} {Phys. Rev. Lett.}\ }\textbf {\bibinfo
  {volume} {119}},\ \bibinfo {pages} {230404} (\bibinfo {year}
  {2017})}\BibitemShut {NoStop}%
\bibitem [{\citenamefont {Wilczek}(2012)}]{Wilczek12}%
  \BibitemOpen
  \bibfield  {author} {\bibinfo {author} {\bibfnamefont {F.}~\bibnamefont
  {Wilczek}},\ }\href {\doibase 10.1103/PhysRevLett.109.160401} {\bibfield
  {journal} {\bibinfo  {journal} {Phys. Rev. Lett.}\ }\textbf {\bibinfo
  {volume} {109}},\ \bibinfo {pages} {160401} (\bibinfo {year}
  {2012})}\BibitemShut {NoStop}%
\bibitem [{\citenamefont {Sacha}(2015)}]{Sacha15}%
  \BibitemOpen
  \bibfield  {author} {\bibinfo {author} {\bibfnamefont {K.}~\bibnamefont
  {Sacha}},\ }\href {\doibase 10.1103/PhysRevA.91.033617} {\bibfield  {journal}
  {\bibinfo  {journal} {Phys. Rev. A}\ }\textbf {\bibinfo {volume} {91}},\
  \bibinfo {pages} {033617} (\bibinfo {year} {2015})}\BibitemShut {NoStop}%
\bibitem [{\citenamefont {Giergiel}\ \emph {et~al.}(2018)\citenamefont
  {Giergiel}, \citenamefont {Miroszewski},\ and\ \citenamefont
  {Sacha}}]{Giergiel18}%
  \BibitemOpen
  \bibfield  {author} {\bibinfo {author} {\bibfnamefont {K.}~\bibnamefont
  {Giergiel}}, \bibinfo {author} {\bibfnamefont {A.}~\bibnamefont
  {Miroszewski}}, \ and\ \bibinfo {author} {\bibfnamefont {K.}~\bibnamefont
  {Sacha}},\ }\href {\doibase 10.1103/PhysRevLett.120.140401} {\bibfield
  {journal} {\bibinfo  {journal} {Phys. Rev. Lett.}\ }\textbf {\bibinfo
  {volume} {120}},\ \bibinfo {pages} {140401} (\bibinfo {year}
  {2018})}\BibitemShut {NoStop}%
\bibitem [{\citenamefont {Sacha}\ and\ \citenamefont
  {Zakrzewski}(2018)}]{Sacha18}%
  \BibitemOpen
  \bibfield  {author} {\bibinfo {author} {\bibfnamefont {K.}~\bibnamefont
  {Sacha}}\ and\ \bibinfo {author} {\bibfnamefont {J.}~\bibnamefont
  {Zakrzewski}},\ }\href {http://stacks.iop.org/0034-4885/81/i=1/a=016401}
  {\bibfield  {journal} {\bibinfo  {journal} {Rep. Prog. Phys.}\ }\textbf
  {\bibinfo {volume} {81}},\ \bibinfo {pages} {016401} (\bibinfo {year}
  {2018})}\BibitemShut {NoStop}%
\bibitem [{\citenamefont {Biddle}\ \emph {et~al.}(2011)\citenamefont {Biddle},
  \citenamefont {Priour}, \citenamefont {Wang},\ and\ \citenamefont
  {Sarma}}]{Biddle11}%
  \BibitemOpen
  \bibfield  {author} {\bibinfo {author} {\bibfnamefont {J.}~\bibnamefont
  {Biddle}}, \bibinfo {author} {\bibfnamefont {D.~J.~J.}\ \bibnamefont
  {Priour}}, \bibinfo {author} {\bibfnamefont {B.}~\bibnamefont {Wang}}, \ and\
  \bibinfo {author} {\bibfnamefont {S.~D.}\ \bibnamefont {Sarma}},\ }\href
  {\doibase 10.1103/PHYSREVB.83.075105} {\bibfield  {journal} {\bibinfo
  {journal} {Physical Review B}\ }\textbf {\bibinfo {volume} {83}},\ \bibinfo
  {pages} {075105} (\bibinfo {year} {2011})}\BibitemShut {NoStop}%
\bibitem [{\citenamefont {Deng}\ \emph {et~al.}(2018)\citenamefont {Deng},
  \citenamefont {Ray}, \citenamefont {Sinha}, \citenamefont {Shlyapnikov},\
  and\ \citenamefont {Santos}}]{Deng18}%
  \BibitemOpen
  \bibfield  {author} {\bibinfo {author} {\bibfnamefont {X.}~\bibnamefont
  {Deng}}, \bibinfo {author} {\bibfnamefont {S.}~\bibnamefont {Ray}}, \bibinfo
  {author} {\bibfnamefont {S.}~\bibnamefont {Sinha}}, \bibinfo {author}
  {\bibfnamefont {G.~S.}\ \bibnamefont {Shlyapnikov}}, \ and\ \bibinfo {author}
  {\bibfnamefont {L.}~\bibnamefont {Santos}},\ }\href@noop {} {\bibfield
  {journal} {\bibinfo  {journal} {preprint arXiv:1808.03585}\ } (\bibinfo
  {year} {2018})}\BibitemShut {NoStop}%
\bibitem [{\citenamefont {L\"uschen}\ \emph {et~al.}(2018)\citenamefont
  {L\"uschen}, \citenamefont {Scherg}, \citenamefont {Kohlert}, \citenamefont
  {Schreiber}, \citenamefont {Bordia}, \citenamefont {Li}, \citenamefont
  {Das~Sarma},\ and\ \citenamefont {Bloch}}]{Lueschen18}%
  \BibitemOpen
  \bibfield  {author} {\bibinfo {author} {\bibfnamefont {H.~P.}\ \bibnamefont
  {L\"uschen}}, \bibinfo {author} {\bibfnamefont {S.}~\bibnamefont {Scherg}},
  \bibinfo {author} {\bibfnamefont {T.}~\bibnamefont {Kohlert}}, \bibinfo
  {author} {\bibfnamefont {M.}~\bibnamefont {Schreiber}}, \bibinfo {author}
  {\bibfnamefont {P.}~\bibnamefont {Bordia}}, \bibinfo {author} {\bibfnamefont
  {X.}~\bibnamefont {Li}}, \bibinfo {author} {\bibfnamefont {S.}~\bibnamefont
  {Das~Sarma}}, \ and\ \bibinfo {author} {\bibfnamefont {I.}~\bibnamefont
  {Bloch}},\ }\href {\doibase 10.1103/PhysRevLett.120.160404} {\bibfield
  {journal} {\bibinfo  {journal} {Phys. Rev. Lett.}\ }\textbf {\bibinfo
  {volume} {120}},\ \bibinfo {pages} {160404} (\bibinfo {year}
  {2018})}\BibitemShut {NoStop}%
\bibitem [{\citenamefont {Flores}(1989)}]{Flores89}%
  \BibitemOpen
  \bibfield  {author} {\bibinfo {author} {\bibfnamefont {J.}~\bibnamefont
  {Flores}},\ }\href {\doibase 10.1088/0953-8984/1/44/017} {\bibfield
  {journal} {\bibinfo  {journal} {J. Phys.: Condens. Matter}\ }\textbf
  {\bibinfo {volume} {1}},\ \bibinfo {pages} {8471} (\bibinfo {year}
  {1989})}\BibitemShut {NoStop}%
\bibitem [{\citenamefont {Izrailev}(2001)}]{Izrailev01}%
  \BibitemOpen
  \bibfield  {author} {\bibinfo {author} {\bibfnamefont {F.}~\bibnamefont
  {Izrailev}},\ }\href {\doibase
  http://dx.doi.org/10.1016/S1386-9477(00)00237-X} {\bibfield  {journal}
  {\bibinfo  {journal} {Physica E: Low-dimensional Systems and Nanostructures}\
  }\textbf {\bibinfo {volume} {9}},\ \bibinfo {pages} {405 } (\bibinfo {year}
  {2001})},\ \bibinfo {note} {proceedings of an International Workshop and
  Seminar on the Dynamics of Complex Systems}\BibitemShut {NoStop}%
\bibitem [{\citenamefont {Peng}\ \emph {et~al.}(2004)\citenamefont {Peng},
  \citenamefont {Liu}, \citenamefont {Huang}, \citenamefont {Qiu},
  \citenamefont {Wang}, \citenamefont {Hu}, \citenamefont {Jiang},
  \citenamefont {Feng}, \citenamefont {Ouyang},\ and\ \citenamefont
  {Zou}}]{Peng04}%
  \BibitemOpen
  \bibfield  {author} {\bibinfo {author} {\bibfnamefont {R.~W.}\ \bibnamefont
  {Peng}}, \bibinfo {author} {\bibfnamefont {Y.~M.}\ \bibnamefont {Liu}},
  \bibinfo {author} {\bibfnamefont {X.~Q.}\ \bibnamefont {Huang}}, \bibinfo
  {author} {\bibfnamefont {F.}~\bibnamefont {Qiu}}, \bibinfo {author}
  {\bibfnamefont {M.}~\bibnamefont {Wang}}, \bibinfo {author} {\bibfnamefont
  {A.}~\bibnamefont {Hu}}, \bibinfo {author} {\bibfnamefont {S.~S.}\
  \bibnamefont {Jiang}}, \bibinfo {author} {\bibfnamefont {D.}~\bibnamefont
  {Feng}}, \bibinfo {author} {\bibfnamefont {L.~Z.}\ \bibnamefont {Ouyang}}, \
  and\ \bibinfo {author} {\bibfnamefont {J.}~\bibnamefont {Zou}},\ }\href
  {\doibase 10.1103/PhysRevB.69.165109} {\bibfield  {journal} {\bibinfo
  {journal} {Phys. Rev. B}\ }\textbf {\bibinfo {volume} {69}},\ \bibinfo
  {pages} {165109} (\bibinfo {year} {2004})}\BibitemShut {NoStop}%
\bibitem [{\citenamefont {Kogan}(2008)}]{Kogan08}%
  \BibitemOpen
  \bibfield  {author} {\bibinfo {author} {\bibfnamefont {E.}~\bibnamefont
  {Kogan}},\ }\href@noop {} {\bibfield  {journal} {\bibinfo  {journal} {Eur.
  Phys. J. B}\ }\textbf {\bibinfo {volume} {61}},\ \bibinfo {pages} {181}
  (\bibinfo {year} {2008})}\BibitemShut {NoStop}%
\bibitem [{\citenamefont {Schaff}\ \emph {et~al.}(2010)\citenamefont {Schaff},
  \citenamefont {Akdeniz},\ and\ \citenamefont {Vignolo}}]{Schaff10}%
  \BibitemOpen
  \bibfield  {author} {\bibinfo {author} {\bibfnamefont {J.-F.}\ \bibnamefont
  {Schaff}}, \bibinfo {author} {\bibfnamefont {Z.}~\bibnamefont {Akdeniz}}, \
  and\ \bibinfo {author} {\bibfnamefont {P.}~\bibnamefont {Vignolo}},\ }\href
  {\doibase 10.1103/PhysRevA.81.041604} {\bibfield  {journal} {\bibinfo
  {journal} {Phys. Rev. A}\ }\textbf {\bibinfo {volume} {81}},\ \bibinfo
  {pages} {041604} (\bibinfo {year} {2010})}\BibitemShut {NoStop}%
\bibitem [{\citenamefont {P\l{}odzie\ifmmode~\acute{n}\else \'{n}\fi{}}\ and\
  \citenamefont {Sacha}(2011)}]{Plodzien11}%
  \BibitemOpen
  \bibfield  {author} {\bibinfo {author} {\bibfnamefont {M.}~\bibnamefont
  {P\l{}odzie\ifmmode~\acute{n}\else \'{n}\fi{}}}\ and\ \bibinfo {author}
  {\bibfnamefont {K.}~\bibnamefont {Sacha}},\ }\href {\doibase
  10.1103/PhysRevA.84.023624} {\bibfield  {journal} {\bibinfo  {journal} {Phys.
  Rev. A}\ }\textbf {\bibinfo {volume} {84}},\ \bibinfo {pages} {023624}
  (\bibinfo {year} {2011})}\BibitemShut {NoStop}%
\bibitem [{\citenamefont {{Wang}}\ \emph {et~al.}(2013)\citenamefont {{Wang}},
  \citenamefont {{Li}},\ and\ \citenamefont {{Nakayama}}}]{Wang13}%
  \BibitemOpen
  \bibfield  {author} {\bibinfo {author} {\bibfnamefont {G.}~\bibnamefont
  {{Wang}}}, \bibinfo {author} {\bibfnamefont {N.}~\bibnamefont {{Li}}}, \ and\
  \bibinfo {author} {\bibfnamefont {T.}~\bibnamefont {{Nakayama}}},\
  }\href@noop {} {\bibfield  {journal} {\bibinfo  {journal} {ArXiv e-prints}\ }
  (\bibinfo {year} {2013})},\ \Eprint {http://arxiv.org/abs/1312.0844}
  {arXiv:1312.0844 [cond-mat.dis-nn]} \BibitemShut {NoStop}%
\bibitem [{\citenamefont {Kosior}\ \emph {et~al.}(2015)\citenamefont {Kosior},
  \citenamefont {Major}, \citenamefont {P\l{}odzie\ifmmode~\acute{n}\else
  \'{n}\fi{}},\ and\ \citenamefont {Zakrzewski}}]{Kosior15}%
  \BibitemOpen
  \bibfield  {author} {\bibinfo {author} {\bibfnamefont {A.}~\bibnamefont
  {Kosior}}, \bibinfo {author} {\bibfnamefont {J.}~\bibnamefont {Major}},
  \bibinfo {author} {\bibfnamefont {M.}~\bibnamefont
  {P\l{}odzie\ifmmode~\acute{n}\else \'{n}\fi{}}}, \ and\ \bibinfo {author}
  {\bibfnamefont {J.}~\bibnamefont {Zakrzewski}},\ }\href {\doibase
  10.1103/PhysRevA.92.023606} {\bibfield  {journal} {\bibinfo  {journal} {Phys.
  Rev. A}\ }\textbf {\bibinfo {volume} {92}},\ \bibinfo {pages} {023606}
  (\bibinfo {year} {2015})}\BibitemShut {NoStop}%
\bibitem [{\citenamefont {{Kosior}}\ \emph {et~al.}(2015)\citenamefont
  {{Kosior}}, \citenamefont {{Major}}, \citenamefont {{P{\l}odzie{\'n}}},\ and\
  \citenamefont {{Zakrzewski}}}]{Kosior15b}%
  \BibitemOpen
  \bibfield  {author} {\bibinfo {author} {\bibfnamefont {A.}~\bibnamefont
  {{Kosior}}}, \bibinfo {author} {\bibfnamefont {J.}~\bibnamefont {{Major}}},
  \bibinfo {author} {\bibfnamefont {M.}~\bibnamefont {{P{\l}odzie{\'n}}}}, \
  and\ \bibinfo {author} {\bibfnamefont {J.}~\bibnamefont {{Zakrzewski}}},\
  }\href {\doibase 10.12693/APhysPolA.128.1002} {\bibfield  {journal} {\bibinfo
   {journal} {Acta. Phys. Pol. A}\ }\textbf {\bibinfo {volume} {128}},\
  \bibinfo {pages} {1002} (\bibinfo {year} {2015})}\BibitemShut {NoStop}%
\bibitem [{\citenamefont {{Liu}}\ \emph {et~al.}(2016)\citenamefont {{Liu}},
  \citenamefont {{Wang}},\ and\ \citenamefont {{Xianlong}}}]{Liu16}%
  \BibitemOpen
  \bibfield  {author} {\bibinfo {author} {\bibfnamefont {T.}~\bibnamefont
  {{Liu}}}, \bibinfo {author} {\bibfnamefont {P.}~\bibnamefont {{Wang}}}, \
  and\ \bibinfo {author} {\bibfnamefont {G.}~\bibnamefont {{Xianlong}}},\
  }\href@noop {} {\bibfield  {journal} {\bibinfo  {journal} {ArXiv e-prints}\ }
  (\bibinfo {year} {2016})},\ \Eprint {http://arxiv.org/abs/1609.06939}
  {arXiv:1609.06939 [cond-mat.dis-nn]} \BibitemShut {NoStop}%
\bibitem [{\citenamefont {Major}(2016)}]{Major16}%
  \BibitemOpen
  \bibfield  {author} {\bibinfo {author} {\bibfnamefont {J.}~\bibnamefont
  {Major}},\ }\href {\doibase 10.1103/PhysRevA.94.053613} {\bibfield  {journal}
  {\bibinfo  {journal} {Phys. Rev. A}\ }\textbf {\bibinfo {volume} {94}},\
  \bibinfo {pages} {053613} (\bibinfo {year} {2016})}\BibitemShut {NoStop}%
\bibitem [{\citenamefont {Larson}\ \emph {et~al.}(2008)\citenamefont {Larson},
  \citenamefont {Damski}, \citenamefont {Morigi},\ and\ \citenamefont
  {Lewenstein}}]{Larson08}%
  \BibitemOpen
  \bibfield  {author} {\bibinfo {author} {\bibfnamefont {J.}~\bibnamefont
  {Larson}}, \bibinfo {author} {\bibfnamefont {B.}~\bibnamefont {Damski}},
  \bibinfo {author} {\bibfnamefont {G.}~\bibnamefont {Morigi}}, \ and\ \bibinfo
  {author} {\bibfnamefont {M.}~\bibnamefont {Lewenstein}},\ }\href {\doibase
  10.1103/PhysRevLett.100.050401} {\bibfield  {journal} {\bibinfo  {journal}
  {Phys. Rev. Lett.}\ }\textbf {\bibinfo {volume} {100}},\ \bibinfo {pages}
  {050401} (\bibinfo {year} {2008})}\BibitemShut {NoStop}%
\bibitem [{\citenamefont {Baumann}\ \emph {et~al.}(2010)\citenamefont
  {Baumann}, \citenamefont {Guerlin}, \citenamefont {Brennecke},\ and\
  \citenamefont {Esslinger}}]{Baumann10}%
  \BibitemOpen
  \bibfield  {author} {\bibinfo {author} {\bibfnamefont {K.}~\bibnamefont
  {Baumann}}, \bibinfo {author} {\bibfnamefont {C.}~\bibnamefont {Guerlin}},
  \bibinfo {author} {\bibfnamefont {F.}~\bibnamefont {Brennecke}}, \ and\
  \bibinfo {author} {\bibfnamefont {T.}~\bibnamefont {Esslinger}},\ }\href
  {http://dx.doi.org/10.1038/nature09009} {\bibfield  {journal} {\bibinfo
  {journal} {Nature}\ }\textbf {\bibinfo {volume} {464}},\ \bibinfo {pages}
  {1301 EP } (\bibinfo {year} {2010})},\ \bibinfo {note} {article}\BibitemShut
  {NoStop}%
\bibitem [{\citenamefont {Fern\'andez-Vidal}\ \emph {et~al.}(2010)\citenamefont
  {Fern\'andez-Vidal}, \citenamefont {De~Chiara}, \citenamefont {Larson},\ and\
  \citenamefont {Morigi}}]{Fernandez10}%
  \BibitemOpen
  \bibfield  {author} {\bibinfo {author} {\bibfnamefont {S.}~\bibnamefont
  {Fern\'andez-Vidal}}, \bibinfo {author} {\bibfnamefont {G.}~\bibnamefont
  {De~Chiara}}, \bibinfo {author} {\bibfnamefont {J.}~\bibnamefont {Larson}}, \
  and\ \bibinfo {author} {\bibfnamefont {G.}~\bibnamefont {Morigi}},\ }\href
  {\doibase 10.1103/PhysRevA.81.043407} {\bibfield  {journal} {\bibinfo
  {journal} {Phys. Rev. A}\ }\textbf {\bibinfo {volume} {81}},\ \bibinfo
  {pages} {043407} (\bibinfo {year} {2010})}\BibitemShut {NoStop}%
\bibitem [{\citenamefont {Ritsch}\ \emph {et~al.}(2013)\citenamefont {Ritsch},
  \citenamefont {Domokos}, \citenamefont {Brennecke},\ and\ \citenamefont
  {Esslinger}}]{Ritsch13}%
  \BibitemOpen
  \bibfield  {author} {\bibinfo {author} {\bibfnamefont {H.}~\bibnamefont
  {Ritsch}}, \bibinfo {author} {\bibfnamefont {P.}~\bibnamefont {Domokos}},
  \bibinfo {author} {\bibfnamefont {F.}~\bibnamefont {Brennecke}}, \ and\
  \bibinfo {author} {\bibfnamefont {T.}~\bibnamefont {Esslinger}},\ }\href
  {\doibase 10.1103/RevModPhys.85.553} {\bibfield  {journal} {\bibinfo
  {journal} {Rev. Mod. Phys.}\ }\textbf {\bibinfo {volume} {85}},\ \bibinfo
  {pages} {553} (\bibinfo {year} {2013})}\BibitemShut {NoStop}%
\bibitem [{\citenamefont {Klinder}\ \emph {et~al.}(2015)\citenamefont
  {Klinder}, \citenamefont {Ke\ss{}ler}, \citenamefont {Bakhtiari},
  \citenamefont {Thorwart},\ and\ \citenamefont {Hemmerich}}]{Klinder15}%
  \BibitemOpen
  \bibfield  {author} {\bibinfo {author} {\bibfnamefont {J.}~\bibnamefont
  {Klinder}}, \bibinfo {author} {\bibfnamefont {H.}~\bibnamefont {Ke\ss{}ler}},
  \bibinfo {author} {\bibfnamefont {M.~R.}\ \bibnamefont {Bakhtiari}}, \bibinfo
  {author} {\bibfnamefont {M.}~\bibnamefont {Thorwart}}, \ and\ \bibinfo
  {author} {\bibfnamefont {A.}~\bibnamefont {Hemmerich}},\ }\href {\doibase
  10.1103/PhysRevLett.115.230403} {\bibfield  {journal} {\bibinfo  {journal}
  {Phys. Rev. Lett.}\ }\textbf {\bibinfo {volume} {115}},\ \bibinfo {pages}
  {230403} (\bibinfo {year} {2015})}\BibitemShut {NoStop}%
\bibitem [{\citenamefont {Landig}\ \emph {et~al.}(2016)\citenamefont {Landig},
  \citenamefont {Hruby}, \citenamefont {Dogra}, \citenamefont {Landini},
  \citenamefont {Mottl}, \citenamefont {Donner},\ and\ \citenamefont
  {Esslinger}}]{Landig16}%
  \BibitemOpen
  \bibfield  {author} {\bibinfo {author} {\bibfnamefont {R.}~\bibnamefont
  {Landig}}, \bibinfo {author} {\bibfnamefont {L.}~\bibnamefont {Hruby}},
  \bibinfo {author} {\bibfnamefont {N.}~\bibnamefont {Dogra}}, \bibinfo
  {author} {\bibfnamefont {M.}~\bibnamefont {Landini}}, \bibinfo {author}
  {\bibfnamefont {R.}~\bibnamefont {Mottl}}, \bibinfo {author} {\bibfnamefont
  {T.}~\bibnamefont {Donner}}, \ and\ \bibinfo {author} {\bibfnamefont
  {T.}~\bibnamefont {Esslinger}},\ }\href
  {http://dx.doi.org/10.1038/nature17409} {\bibfield  {journal} {\bibinfo
  {journal} {Nature}\ }\textbf {\bibinfo {volume} {532}},\ \bibinfo {pages}
  {476 EP } (\bibinfo {year} {2016})}\BibitemShut {NoStop}%
\bibitem [{\citenamefont {Dogra}\ \emph {et~al.}(2016)\citenamefont {Dogra},
  \citenamefont {Brennecke}, \citenamefont {Huber},\ and\ \citenamefont
  {Donner}}]{Dogra16}%
  \BibitemOpen
  \bibfield  {author} {\bibinfo {author} {\bibfnamefont {N.}~\bibnamefont
  {Dogra}}, \bibinfo {author} {\bibfnamefont {F.}~\bibnamefont {Brennecke}},
  \bibinfo {author} {\bibfnamefont {S.~D.}\ \bibnamefont {Huber}}, \ and\
  \bibinfo {author} {\bibfnamefont {T.}~\bibnamefont {Donner}},\ }\href
  {\doibase 10.1103/PhysRevA.94.023632} {\bibfield  {journal} {\bibinfo
  {journal} {Phys. Rev. A}\ }\textbf {\bibinfo {volume} {94}},\ \bibinfo
  {pages} {023632} (\bibinfo {year} {2016})}\BibitemShut {NoStop}%
\bibitem [{\citenamefont {Niederle}\ \emph {et~al.}(2016)\citenamefont
  {Niederle}, \citenamefont {Morigi},\ and\ \citenamefont
  {Rieger}}]{Niederle16}%
  \BibitemOpen
  \bibfield  {author} {\bibinfo {author} {\bibfnamefont {A.~E.}\ \bibnamefont
  {Niederle}}, \bibinfo {author} {\bibfnamefont {G.}~\bibnamefont {Morigi}}, \
  and\ \bibinfo {author} {\bibfnamefont {H.}~\bibnamefont {Rieger}},\ }\href
  {\doibase 10.1103/PhysRevA.94.033607} {\bibfield  {journal} {\bibinfo
  {journal} {Phys. Rev. A}\ }\textbf {\bibinfo {volume} {94}},\ \bibinfo
  {pages} {033607} (\bibinfo {year} {2016})}\BibitemShut {NoStop}%
\bibitem [{\citenamefont {Habibian}\ \emph
  {et~al.}(2013{\natexlab{a}})\citenamefont {Habibian}, \citenamefont {Winter},
  \citenamefont {Paganelli}, \citenamefont {Rieger},\ and\ \citenamefont
  {Morigi}}]{Habibian13}%
  \BibitemOpen
  \bibfield  {author} {\bibinfo {author} {\bibfnamefont {H.}~\bibnamefont
  {Habibian}}, \bibinfo {author} {\bibfnamefont {A.}~\bibnamefont {Winter}},
  \bibinfo {author} {\bibfnamefont {S.}~\bibnamefont {Paganelli}}, \bibinfo
  {author} {\bibfnamefont {H.}~\bibnamefont {Rieger}}, \ and\ \bibinfo {author}
  {\bibfnamefont {G.}~\bibnamefont {Morigi}},\ }\href {\doibase
  10.1103/PhysRevLett.110.075304} {\bibfield  {journal} {\bibinfo  {journal}
  {Phys. Rev. Lett.}\ }\textbf {\bibinfo {volume} {110}},\ \bibinfo {pages}
  {075304} (\bibinfo {year} {2013}{\natexlab{a}})}\BibitemShut {NoStop}%
\bibitem [{\citenamefont {Habibian}\ \emph
  {et~al.}(2013{\natexlab{b}})\citenamefont {Habibian}, \citenamefont {Winter},
  \citenamefont {Paganelli}, \citenamefont {Rieger},\ and\ \citenamefont
  {Morigi}}]{Habibian13b}%
  \BibitemOpen
  \bibfield  {author} {\bibinfo {author} {\bibfnamefont {H.}~\bibnamefont
  {Habibian}}, \bibinfo {author} {\bibfnamefont {A.}~\bibnamefont {Winter}},
  \bibinfo {author} {\bibfnamefont {S.}~\bibnamefont {Paganelli}}, \bibinfo
  {author} {\bibfnamefont {H.}~\bibnamefont {Rieger}}, \ and\ \bibinfo {author}
  {\bibfnamefont {G.}~\bibnamefont {Morigi}},\ }\href {\doibase
  10.1103/PhysRevA.88.043618} {\bibfield  {journal} {\bibinfo  {journal} {Phys.
  Rev. A}\ }\textbf {\bibinfo {volume} {88}},\ \bibinfo {pages} {043618}
  (\bibinfo {year} {2013}{\natexlab{b}})}\BibitemShut {NoStop}%
\bibitem [{\citenamefont {Rojan}\ \emph {et~al.}(2016)\citenamefont {Rojan},
  \citenamefont {Kraus}, \citenamefont {Fogarty}, \citenamefont {Habibian},
  \citenamefont {Minguzzi},\ and\ \citenamefont {Morigi}}]{Rojan16}%
  \BibitemOpen
  \bibfield  {author} {\bibinfo {author} {\bibfnamefont {K.}~\bibnamefont
  {Rojan}}, \bibinfo {author} {\bibfnamefont {R.}~\bibnamefont {Kraus}},
  \bibinfo {author} {\bibfnamefont {T.}~\bibnamefont {Fogarty}}, \bibinfo
  {author} {\bibfnamefont {H.}~\bibnamefont {Habibian}}, \bibinfo {author}
  {\bibfnamefont {A.}~\bibnamefont {Minguzzi}}, \ and\ \bibinfo {author}
  {\bibfnamefont {G.}~\bibnamefont {Morigi}},\ }\href {\doibase
  10.1103/PhysRevA.94.013839} {\bibfield  {journal} {\bibinfo  {journal} {Phys.
  Rev. A}\ }\textbf {\bibinfo {volume} {94}},\ \bibinfo {pages} {013839}
  (\bibinfo {year} {2016})}\BibitemShut {NoStop}%
\bibitem [{\citenamefont {Zheng}\ and\ \citenamefont {Cooper}(2018)}]{Zheng18}%
  \BibitemOpen
  \bibfield  {author} {\bibinfo {author} {\bibfnamefont {W.}~\bibnamefont
  {Zheng}}\ and\ \bibinfo {author} {\bibfnamefont {N.~R.}\ \bibnamefont
  {Cooper}},\ }\href {\doibase 10.1103/PhysRevA.97.021601} {\bibfield
  {journal} {\bibinfo  {journal} {Phys. Rev. A}\ }\textbf {\bibinfo {volume}
  {97}},\ \bibinfo {pages} {021601} (\bibinfo {year} {2018})}\BibitemShut
  {NoStop}%
\bibitem [{\citenamefont {Kimble}(9944)}]{Kimble94}%
  \BibitemOpen
  \bibfield  {author} {\bibinfo {author} {\bibfnamefont {J.}~\bibnamefont
  {Kimble}},\ }in\ \href@noop {} {\emph {\bibinfo {booktitle} {Cavity Quantum
  Electrodynamics}}},\ \bibinfo {editor} {edited by\ \bibinfo {editor}
  {\bibfnamefont {P.~R.}\ \bibnamefont {Berman}}}\ (\bibinfo  {publisher}
  {Academic},\ \bibinfo {address} {New York},\ \bibinfo {year} {19944})\ p.\
  \bibinfo {pages} {203}\BibitemShut {NoStop}%
\bibitem [{\citenamefont {Sch\"utz}\ \emph {et~al.}(2013)\citenamefont
  {Sch\"utz}, \citenamefont {Habibian},\ and\ \citenamefont
  {Morigi}}]{Schuetz13}%
  \BibitemOpen
  \bibfield  {author} {\bibinfo {author} {\bibfnamefont {S.}~\bibnamefont
  {Sch\"utz}}, \bibinfo {author} {\bibfnamefont {H.}~\bibnamefont {Habibian}},
  \ and\ \bibinfo {author} {\bibfnamefont {G.}~\bibnamefont {Morigi}},\ }\href
  {\doibase 10.1103/PhysRevA.88.033427} {\bibfield  {journal} {\bibinfo
  {journal} {Phys. Rev. A}\ }\textbf {\bibinfo {volume} {88}},\ \bibinfo
  {pages} {033427} (\bibinfo {year} {2013})}\BibitemShut {NoStop}%
\bibitem [{\citenamefont {Miller}\ \emph {et~al.}(2005)\citenamefont {Miller},
  \citenamefont {Northup}, \citenamefont {Birnbaum}, \citenamefont {Boca},
  \citenamefont {Boozer},\ and\ \citenamefont {Kimble}}]{Miller05}%
  \BibitemOpen
  \bibfield  {author} {\bibinfo {author} {\bibfnamefont {R.}~\bibnamefont
  {Miller}}, \bibinfo {author} {\bibfnamefont {T.~E.}\ \bibnamefont {Northup}},
  \bibinfo {author} {\bibfnamefont {K.~M.}\ \bibnamefont {Birnbaum}}, \bibinfo
  {author} {\bibfnamefont {A.}~\bibnamefont {Boca}}, \bibinfo {author}
  {\bibfnamefont {A.~D.}\ \bibnamefont {Boozer}}, \ and\ \bibinfo {author}
  {\bibfnamefont {H.~J.}\ \bibnamefont {Kimble}},\ }\href
  {http://stacks.iop.org/0953-4075/38/i=9/a=007} {\bibfield  {journal}
  {\bibinfo  {journal} {Journal of Physics B: Atomic, Molecular and Optical
  Physics}\ }\textbf {\bibinfo {volume} {38}},\ \bibinfo {pages} {S551}
  (\bibinfo {year} {2005})}\BibitemShut {NoStop}%
\bibitem [{\citenamefont {Reimann}\ \emph {et~al.}(2015)\citenamefont
  {Reimann}, \citenamefont {Alt}, \citenamefont {Kampschulte}, \citenamefont
  {Macha}, \citenamefont {Ratschbacher}, \citenamefont {Thau}, \citenamefont
  {Yoon},\ and\ \citenamefont {Meschede}}]{Reimann15}%
  \BibitemOpen
  \bibfield  {author} {\bibinfo {author} {\bibfnamefont {R.}~\bibnamefont
  {Reimann}}, \bibinfo {author} {\bibfnamefont {W.}~\bibnamefont {Alt}},
  \bibinfo {author} {\bibfnamefont {T.}~\bibnamefont {Kampschulte}}, \bibinfo
  {author} {\bibfnamefont {T.}~\bibnamefont {Macha}}, \bibinfo {author}
  {\bibfnamefont {L.}~\bibnamefont {Ratschbacher}}, \bibinfo {author}
  {\bibfnamefont {N.}~\bibnamefont {Thau}}, \bibinfo {author} {\bibfnamefont
  {S.}~\bibnamefont {Yoon}}, \ and\ \bibinfo {author} {\bibfnamefont
  {D.}~\bibnamefont {Meschede}},\ }\href {\doibase
  10.1103/PhysRevLett.114.023601} {\bibfield  {journal} {\bibinfo  {journal}
  {Phys. Rev. Lett.}\ }\textbf {\bibinfo {volume} {114}},\ \bibinfo {pages}
  {023601} (\bibinfo {year} {2015})}\BibitemShut {NoStop}%
\bibitem [{\citenamefont {Neuzner}\ \emph {et~al.}(2016)\citenamefont
  {Neuzner}, \citenamefont {K\"orber}, \citenamefont {Morin}, \citenamefont
  {Ritter},\ and\ \citenamefont {Rempe}}]{Neuzner16}%
  \BibitemOpen
  \bibfield  {author} {\bibinfo {author} {\bibfnamefont {A.}~\bibnamefont
  {Neuzner}}, \bibinfo {author} {\bibfnamefont {M.}~\bibnamefont {K\"orber}},
  \bibinfo {author} {\bibfnamefont {O.}~\bibnamefont {Morin}}, \bibinfo
  {author} {\bibfnamefont {S.}~\bibnamefont {Ritter}}, \ and\ \bibinfo {author}
  {\bibfnamefont {G.}~\bibnamefont {Rempe}},\ }\href {\doibase
  10.1038/nphoton.2016.19} {\bibfield  {journal} {\bibinfo  {journal} {Nature
  Photonics}\ }\textbf {\bibinfo {volume} {10}},\ \bibinfo {pages} {303}
  (\bibinfo {year} {2016})}\BibitemShut {NoStop}%
\bibitem [{\citenamefont {Mott}(1987)}]{Mott87}%
  \BibitemOpen
  \bibfield  {author} {\bibinfo {author} {\bibfnamefont {N.}~\bibnamefont
  {Mott}},\ }\href {http://stacks.iop.org/0022-3719/20/i=21/a=008} {\bibfield
  {journal} {\bibinfo  {journal} {Journal of Physics C: Solid State Physics}\
  }\textbf {\bibinfo {volume} {20}},\ \bibinfo {pages} {3075} (\bibinfo {year}
  {1987})}\BibitemShut {NoStop}%
\bibitem [{\citenamefont {Thouless}(1974)}]{Thouless74}%
  \BibitemOpen
  \bibfield  {author} {\bibinfo {author} {\bibfnamefont {D.~J.}\ \bibnamefont
  {Thouless}},\ }\href@noop {} {\bibfield  {journal} {\bibinfo  {journal}
  {Phys. Rep.}\ }\textbf {\bibinfo {volume} {13}},\ \bibinfo {pages} {93}
  (\bibinfo {year} {1974})}\BibitemShut {NoStop}%
\bibitem [{\citenamefont {Kramer}(1993)}]{Kramer93}%
  \BibitemOpen
  \bibfield  {author} {\bibinfo {author} {\bibfnamefont {A.}~\bibnamefont
  {Kramer}, \bibfnamefont {B.;~MacKinnon}},\ }\href@noop {} {\bibfield
  {journal} {\bibinfo  {journal} {Rep. Prog. Phys.}\ }\textbf {\bibinfo
  {volume} {56}},\ \bibinfo {pages} {1469} (\bibinfo {year}
  {1993})}\BibitemShut {NoStop}%
\bibitem [{\citenamefont {Fogarty}\ \emph {et~al.}(2015)\citenamefont
  {Fogarty}, \citenamefont {Cormick}, \citenamefont {Landa}, \citenamefont
  {Stojanovi\ifmmode~\acute{c}\else \'{c}\fi{}}, \citenamefont {Demler},\ and\
  \citenamefont {Morigi}}]{Fogarty15}%
  \BibitemOpen
  \bibfield  {author} {\bibinfo {author} {\bibfnamefont {T.}~\bibnamefont
  {Fogarty}}, \bibinfo {author} {\bibfnamefont {C.}~\bibnamefont {Cormick}},
  \bibinfo {author} {\bibfnamefont {H.}~\bibnamefont {Landa}}, \bibinfo
  {author} {\bibfnamefont {V.~M.}\ \bibnamefont
  {Stojanovi\ifmmode~\acute{c}\else \'{c}\fi{}}}, \bibinfo {author}
  {\bibfnamefont {E.}~\bibnamefont {Demler}}, \ and\ \bibinfo {author}
  {\bibfnamefont {G.}~\bibnamefont {Morigi}},\ }\href {\doibase
  10.1103/PhysRevLett.115.233602} {\bibfield  {journal} {\bibinfo  {journal}
  {Phys. Rev. Lett.}\ }\textbf {\bibinfo {volume} {115}},\ \bibinfo {pages}
  {233602} (\bibinfo {year} {2015})}\BibitemShut {NoStop}%
\bibitem [{\citenamefont {Hurwitz}(1891)}]{Hurwitz91}%
  \BibitemOpen
  \bibfield  {author} {\bibinfo {author} {\bibfnamefont {A.}~\bibnamefont
  {Hurwitz}},\ }\href {\doibase 10.1007/BF01206656} {\bibfield  {journal}
  {\bibinfo  {journal} {Mathematische Annalen}\ }\textbf {\bibinfo {volume}
  {39}},\ \bibinfo {pages} {279} (\bibinfo {year} {1891})}\BibitemShut
  {NoStop}%
\bibitem [{\citenamefont {Prasanna~Venkatesh}\ \emph
  {et~al.}(2009)\citenamefont {Prasanna~Venkatesh}, \citenamefont {Trupke},
  \citenamefont {Hinds},\ and\ \citenamefont {O'Dell}}]{Venkatesh09}%
  \BibitemOpen
  \bibfield  {author} {\bibinfo {author} {\bibfnamefont {B.}~\bibnamefont
  {Prasanna~Venkatesh}}, \bibinfo {author} {\bibfnamefont {M.}~\bibnamefont
  {Trupke}}, \bibinfo {author} {\bibfnamefont {E.~A.}\ \bibnamefont {Hinds}}, \
  and\ \bibinfo {author} {\bibfnamefont {D.~H.~J.}\ \bibnamefont {O'Dell}},\
  }\href {\doibase 10.1103/PhysRevA.80.063834} {\bibfield  {journal} {\bibinfo
  {journal} {Phys. Rev. A}\ }\textbf {\bibinfo {volume} {80}},\ \bibinfo
  {pages} {063834} (\bibinfo {year} {2009})}\BibitemShut {NoStop}%
\bibitem [{\citenamefont {Venkatesh}\ and\ \citenamefont
  {O'Dell}(2013)}]{Venkatesh13}%
  \BibitemOpen
  \bibfield  {author} {\bibinfo {author} {\bibfnamefont {B.~P.}\ \bibnamefont
  {Venkatesh}}\ and\ \bibinfo {author} {\bibfnamefont {D.~H.~J.}\ \bibnamefont
  {O'Dell}},\ }\href {\doibase 10.1103/PhysRevA.88.013848} {\bibfield
  {journal} {\bibinfo  {journal} {Phys. Rev. A}\ }\textbf {\bibinfo {volume}
  {88}},\ \bibinfo {pages} {013848} (\bibinfo {year} {2013})}\BibitemShut
  {NoStop}%
\bibitem [{\citenamefont {Georges}\ \emph {et~al.}(2017)\citenamefont
  {Georges}, \citenamefont {Vargas}, \citenamefont {Ke\ss{}ler}, \citenamefont
  {Klinder},\ and\ \citenamefont {Hemmerich}}]{Georges17}%
  \BibitemOpen
  \bibfield  {author} {\bibinfo {author} {\bibfnamefont {C.}~\bibnamefont
  {Georges}}, \bibinfo {author} {\bibfnamefont {J.}~\bibnamefont {Vargas}},
  \bibinfo {author} {\bibfnamefont {H.}~\bibnamefont {Ke\ss{}ler}}, \bibinfo
  {author} {\bibfnamefont {J.}~\bibnamefont {Klinder}}, \ and\ \bibinfo
  {author} {\bibfnamefont {A.}~\bibnamefont {Hemmerich}},\ }\href {\doibase
  10.1103/PhysRevA.96.063615} {\bibfield  {journal} {\bibinfo  {journal} {Phys.
  Rev. A}\ }\textbf {\bibinfo {volume} {96}},\ \bibinfo {pages} {063615}
  (\bibinfo {year} {2017})}\BibitemShut {NoStop}%
\bibitem [{\citenamefont {Li}\ and\ \citenamefont {Das~Sarma}(2018)}]{Li18}%
  \BibitemOpen
  \bibfield  {author} {\bibinfo {author} {\bibfnamefont {X.}~\bibnamefont
  {Li}}\ and\ \bibinfo {author} {\bibfnamefont {S.}~\bibnamefont {Das~Sarma}},\
  }\href@noop {} {\bibfield  {journal} {\bibinfo  {journal} {preprint
  arXiv:1808.05954}\ } (\bibinfo {year} {2018})}\BibitemShut {NoStop}%
\end{thebibliography}
%


\end{document}